\documentclass{aastex631}

\shorttitle{Blazar variability with Rubin-LSST}
\shortauthors{Raiteri et al.}
\graphicspath{{./}{figures/}}

\begin{document}

\title{Blazar variability with the Vera C. Rubin Legacy Survey of Space and Time (LSST).}

\correspondingauthor{Claudia M. Raiteri}
\email{claudia.raiteri@inaf.it}

\author[0000-0003-1784-2784]{Claudia M. Raiteri}
\affiliation{INAF-Osservatorio Astrofisico di Torino,
Via Osservatorio 20, I-10025 Pino Torinese, Italy}

\author[0000-0001-5843-5515]{Maria I. Carnerero}
\affiliation{INAF-Osservatorio Astrofisico di Torino,
Via Osservatorio 20, I-10025 Pino Torinese, Italy}

\author[0000-0002-0690-0638]{Barbara Balmaverde}
\affiliation{INAF-Osservatorio Astrofisico di Torino,
Via Osservatorio 20, I-10025 Pino Torinese, Italy}

\author[0000-0001-8018-5348]{Eric C. Bellm}
\affiliation{DIRAC Institute, Department of Astronomy, University of Washington, 3910 15th Avenue NE, Seattle, WA 98195, USA}

\author[0000-0002-2577-8885]{William Clarkson}
\affiliation{University of Michigan-Dearborn,
Dearborn, MI, US}

\author[0000-0001-7618-7527]{Filippo D'Ammando}
\affiliation{INAF-Istituto di Radioastronomia,
Via Gobetti 101, I-40129 Bologna, Italy}

\author[0000-0003-4210-7693]{Maurizio Paolillo}
\affiliation{Università degli Studi di Napoli Federico II, I-80126 Napoli, Italy}

\author[0000-0002-1061-1804]{Gordon T. Richards}
\affiliation{Department of Physics, Drexel University, 32 S. 32nd Street, Philadelphia, PA 19104, USA}

\author[0000-0003-1743-6946]{Massimo Villata}
\affiliation{INAF-Osservatorio Astrofisico di Torino,
Via Osservatorio 20, I-10025 Pino Torinese, Italy}

\author[0000-0003-2874-6464]{Peter Yoachim}
\affiliation{University of Washington, Seattle, WA, US}

\author[0000-0001-9163-0064]{Ilsang Yoon}
\affiliation{National Radio Astronomy Observatory, Charlottesville, VA, US}



\begin{abstract}
With their emission mainly coming from a relativistic jet pointing towards us, blazars are fundamental sources to study extragalactic jets and their central engines, consisting of supermassive black holes (SMBHs) fed by accretion discs. 
They are also candidate sources of high-energy neutrinos and cosmic rays.
Because of the jet orientation, the non-thermal blazar emission is Doppler beamed; its variability is unpredictable and occurs on time-scales from less than one hour to years. 
The comprehension of the diverse mechanisms producing the flux and spectral changes requires well-sampled multiband light curves on long time periods. 
In particular, outbursts are the best test bench to shed light on the underlying physics, especially when studied in a multiwavelength context.
The Vera C. Rubin Legacy Survey of Space and Time (Rubin-LSST) will monitor the southern sky for ten years in six photometric bands, offering a formidable tool to study blazar variability features in a statistical way. 
The alert system will allow us to trigger follow-up observations of outstanding events, especially at high (keV-to-GeV) and very high (TeV) energies. 
We here examine the simulated Rubin-LSST survey strategies with the aim of understanding which cadences are more suitable for the blazar variability science.
Our metrics include light curve and colour sampling. 
We also investigate the problem of saturation, which will affect the brightest and many flaring sources, and will have a detrimental impact on follow-up observations.

\end{abstract}

\keywords{Active galactic nuclei (16) --- Blazars (164) --- Flat-spectrum radio quasars (2163) --- BL Lacertae objects (158) --- Plasma jets (1263)}

\section{Introduction} \label{intro}
At the centre of many galaxies, supermassive black holes (SMBHs) of millions to billions solar masses are fed by matter falling from a surrounding accretion disc, releasing gravitational energy. These active galactic nuclei (AGN) show a variety of different properties, also depending on their orientation with respect to the line of sight \citep[e.g.][]{urry1995}. In radio-loud objects, two plasma jets are emitted (nearly) perpendicularly to the disc. When one of the jets is oriented closely to the line of sight, its emission is relativistically Doppler beamed and usually dominates over all the other sources of radiation. The objects where this occurs are called ``blazars''. They show random flux variability at all wavelengths, from the radio band to the $\gamma$ rays, with time-scales ranging from less than one hour to years. In general, low-amplitude fast variations overlap with larger and slower flux oscillations, indicating that different variability mechanisms are at work. Spectral variability is also usually detected, with sometimes definite trends with brightness.

The Vera C. Rubin Legacy Survey of Space and Time (Rubin-LSST) offers a magnificent opportunity to study the whole population of blazars in the southern hemisphere and beyond, and to address the still open questions about their variability, census and environment.
The 8.4 m telescope in Chile will scan more than 18 000 square degrees of sky for ten years in the six bands $u, g, r, i, z, y$ \citep{ivezic2019}. The field of view will be 9.6 square degrees. At least 80\% of the Rubin-LSST time will be dedicated to the Main Survey, the Wide-Fast-Deep (WFD). Four fields will be observed at a much higher rate: these deep drilling fields (DDFs) will cover regions of the sky where an intensive multiwavelength observing effort has already been spent. A further double DDF will match the Euclid Deep Field South\footnote{https://www.cosmos.esa.int/web/euclid/euclid-survey}. Moreover, three minisurveys will likely explore the Galactic Plane, The North Ecliptic Spur and the South Celestial Pole, though with a lower cadence than the WFD.

Through the study of blazar variability, we aim at understanding what happens in extragalactic jets at (sub)parsec scales. There, particles are accelerated at relativistic speeds, emitting synchrotron radiation observed in the radio-to-UV\footnote{In some objects up to the X-ray band.} frequency bands, and X-ray and $\gamma$-ray photons through inverse-Compton and/or hadronic processes \citep[e.g.][]{boettcher2013}. Blazars are formidable cosmic accelerators, which are also alleged to produce high-energy neutrinos and cosmic rays \citep[e.g.][]{murase2012,ice2018a,ice2018b,giommi2020}. 
Moreover, through the analysis of blazar variability we can shed light on the jet structure and its time evolution.
The variability mechanisms can be both intrinsic and extrinsic. The intrinsic ones are probably due to particle injection or acceleration in the jet, possibly produced by shock waves propagating downstream, or to magnetic reconnection \citep[e.g.][]{boettcher2019,bodo2021}. The extrinsic mechanisms have a geometrical nature, caused by changes in the orientation of the jet emitting regions with respect to the line of sight, which produces variations of the corresponding Doppler beaming factor \citep[e.g.][]{villata1999,raiteri2017}. Both intrinsic and extrinsic mechanisms may be at work on different time-scales.

If multiband optical light curves are available, as in Rubin-LSST, we can follow the spectral changes by means of colour indices, provided these are obtained with data taken close in time. In case of nearly contemporaneous data at several different frequencies in the optical band and beyond, we can build broad-band spectral energy distributions (SEDs).
The shape of the broad-band SED can identify the blazar type [low-~, intermediate-~, or high-energy peaked, depending on the frequency $\nu$ of the synchrotron peak in the $\log (\nu \, F_\nu)$ versus $\log \nu$ representation, where $F_\nu$ is the flux density]. Some sources have been seen to change type according to their brightness level.
The spectral behaviour in time depends on the variability mechanism: achromatism or quasi-achromatism is compatible with a geometrical interpretation, while strong chromatism implies energetic processes.
In some cases, spectral changes are found much stronger on short than on long time-scales. This has been interpreted as the result of rapid, intrinsic variability in jet regions that undergo orientation changes \citep{raiteri2021a,raiteri2021b}. 

In this paper, we will investigate the impact of different choices for the Rubin-LSST observing strategy on blazar variability studies. 
We will make use of the well-sampled and carefully checked blazar light curves of the Whole Earth Blazar Telescope\footnote{http://www.oato.inaf.it/blazars/webt} \citep[WEBT, e.g.][]{villata2002,raiteri2017}.
Born in 1997, the WEBT is an international collaboration of many tens of astronomers, who monitor a number of sources mainly in the optical bands, but also at radio and less frequently at near-infrared wavelengths. 
We will adopt the Roma-BZCAT5 \citep{massaro2009,massaro2015} as the reference blazar catalogue, since it includes confirmed sources and an estimate of their optical brightness.
However, we will also discuss the implications of other blazar catalogues, containing tens of thousands candidates.

\section{Emission contributions in the optical band}
The blazar AGN class includes flat-spectrum radio quasars (FSRQs) and BL Lac-type objects. They are classically distinguished on the base of the strength of their emission lines: BL Lac-type sources show featureless spectra or lines with equivalent width smaller than 5 \AA\ in the rest frame \citep[][]{stickel1991,stocke1991}. This definition is somewhat problematic, because it depends on the relative importance of the featureless jet emission contribution against that coming from the nuclear region, which includes the accretion disc and the broad- and narrow-line regions. Even the prototype of BL Lac-type objects, BL Lacertae, does not behave as a BL Lac-type source when its synchrotron emission is weak \citep[][]{vermeulen1995,corbett2000}.
While in the radio or high-energy bands the beamed non-thermal radiation from the jet dominates, in the optical the scenario is more complex, and we can in general meet three different situations, which are described below.

\subsection{Host-dominated BL Lac-type objects}
The optical emission of close BL Lac-type sources can be dominated by the light of the host galaxy. In BZCAT5 these objects represent $\sim 7.7\%$ of the whole blazar population and the most distant one is 5BZGJ1552+3159 at redshift $z=0.584$.
These low-luminosity sources are recognized as blazars because of their multiwavelength behaviour and/or because of anomalous Ca break value \citep[e.g.][]{capetti2015}.
The study of the flux and spectral variability of the jet component requires the determination of the host galaxy contribution to the total emission in the various bands, in order to subtract the host galaxy light from the total flux.

\subsection{Disc-dominated FSRQs}
When the thermal radiation due to the accretion disc (the ``big blue bump") prevails against the non-thermal jet emission, these objects behave like normal quasars and their variability is well represented as a damped random walk \citep[e.g.][]{macleod2010,butler2011}. 
If a strong non-thermal flare occurs, it can be distinguished from the disc variability as a more rapid event which is likely seen also outside the optical band, particularly at $\gamma$ rays. 
 
The spectra of these objects show strong broad emission lines that can be detected as excess in some specific Rubin-LSST bands, which depend on the redshift, and can likely allow photometric reverberation mapping studies \citep[e.g.][]{chelouche2012}.

\subsection{Synchrotron-dominated sources}
The most interesting sources from the point of view of blazar variability and jet understanding are BL Lac-type objects and FSRQs where the non-thermal radiation is playing the major role. The observed emission is subject to the rules of relativistic beaming, with Doppler-enhanced flux and variability amplitudes, and Doppler-shortened time-scales.
The most extreme objects are called optically violent variable (OVV) sources.
Long-term and well-sampled optical light curves in different bands are needed to cover the variety of variability time-scales. 
Spectral variability prevents the use of data in different filters to increase the sampling of a given light curve: if average colour indices are adopted, the corresponding errors may reach several tenths of mag. 

These sources can undergo major outbursts, with brightness increase up to several magnitudes \citep[e.g.][]{raiteri2017}, so that saturation in Rubin-LSST data will affect some of the most critical and interesting events. In the following sections we will also address this problem.

\section{Rubin-LSST observing strategies}
\label{sec:cadences}
The observing strategy of Rubin-LSST must meet the original goals \citep{ivezic2019} and at the same time optimize as many science cases as possible. 
Many cadence simulations (\texttt{OpSim} runs) have been performed
to allow researchers in all astronomy fields to test their science cases through the Metric Analysis Framework \citep[\texttt{MAF};][]{jones2014} software.
These simulations cover many different possibilities, like changing the survey footprint or the filter distribution, adding a third visit to the standard two visits of each field in the same night, considering rolling cadences with some regions of the sky better sampled for certain periods of time, adding short exposures a few times per year.
The international astronomical community is called to provide metrics that can test the performance of the various \texttt{OpSim} runs for their scientific goals \citep{bianco2021}.
A detailed description and analysis of the cadence simulations can be found in 
\citet{jones2020}\footnote{The list of the \texttt{OpSim} runs included in the releases v1.5, v1.6, and v1.7 is available at http://astro-lsst-01.astro.washington.edu:8081, where the results of the application of standard metrics are also given.}.

In this paper, we will use metrics to explore how different \texttt{OpSim} runs impact on the investigation of blazar variability.
We will adopt the \texttt{OpSim} run \texttt{baseline\textunderscore nexp2\textunderscore v1.7\textunderscore 10yrs} as the reference cadence simulation.
It includes two consecutive snapshots with exposure time of 15 s each, for each of the two visits of a given field in the same night.

\section{How many blazars?}
\label{sec:census}

\begin{figure}[ht!]
\plotone{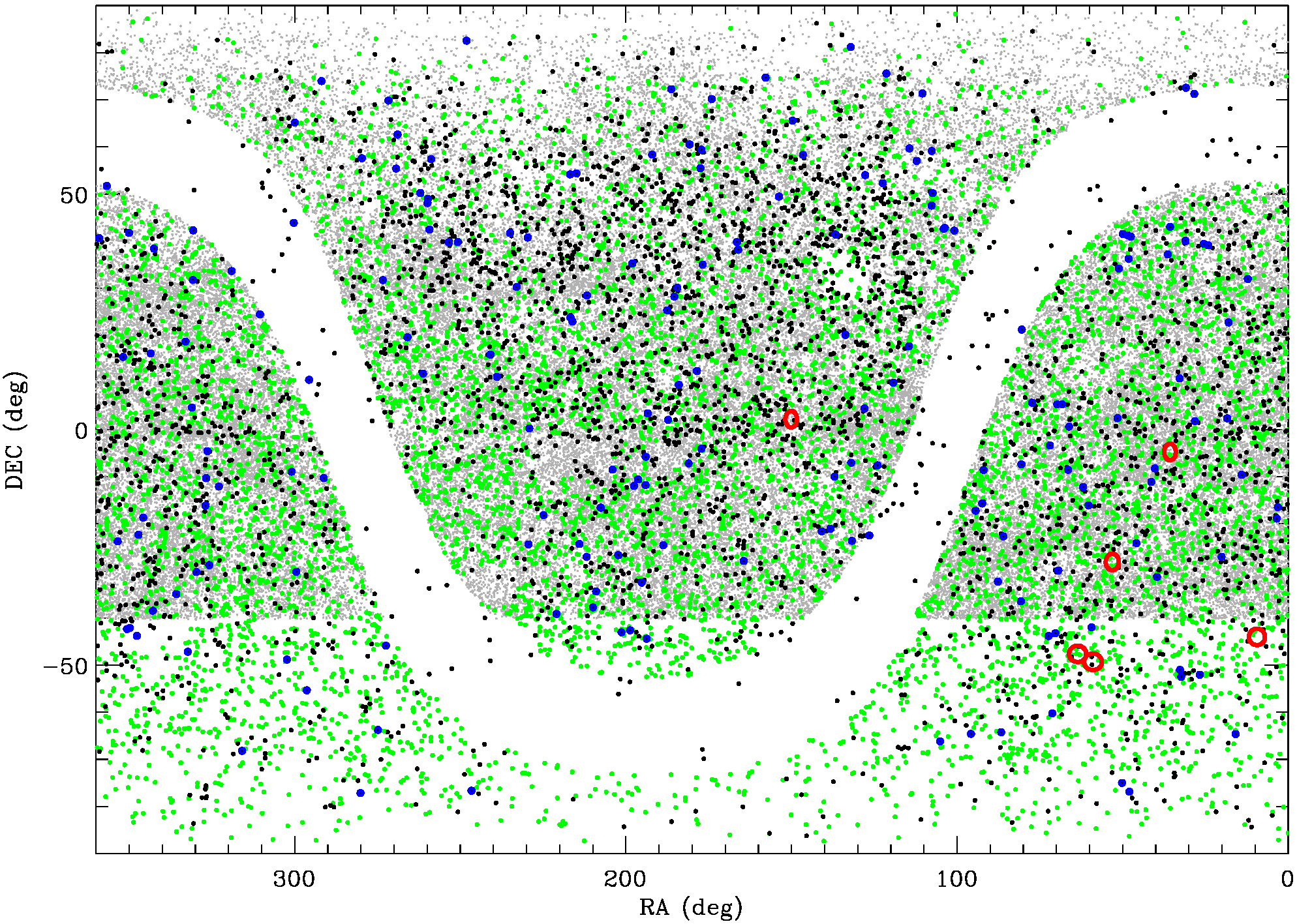}
\caption{Map of the sky with the blazars in the BZCAT5, BROS, and CRATES catalogues plotted with black, grey, and green dots, respectively. BZCAT5 objects with $R$-band catalogue magnitude brighter than 15.5, which is the median maximum saturation magnitude in the LSST $r$ band, are highlighted with blue circles. The regions enclosed by red lines represent the planned DDFs; from North to South: COSMOS, XMM-LSS, ECDFS, ELAIS S1, and EDFS (double field).  \label{fig:mappa}}
\end{figure}

Figure \ref{fig:mappa} shows the distribution in the sky of the 3561 blazars listed in the BZCAT5 catalogue\footnote{https://www.asdc.asi.it/bzcat \citep[][]{massaro2009,massaro2015}}. 
The objects of BZCAT5 were collected from the literature and selected in different ways. The catalogue gives multifrequency information, including optical magnitudes in the $R$ band down to 24.4 mag. 
In the figure, the location of the selected DDFs is indicated, according to \citet{jones2020}. In order of decreasing declination we find: COSMOS, XMM-LSS, ECDFS, ELAIS S1, EDFS (the double field to match to Euclid Deep Field South).
Three DDFs do not contain BZCAT5 blazars. The other three include only five sources overall.

Were BZCAT5 a complete catalogue, DDFs would be of no interest for blazar studies. However, there are other catalogues of blazar candidates, comprising much more objects. 
The CRATES catalogue \citep{healey2007} lists about $11 \, 000$ flat-spectrum radio sources out of the Galactic Plane. The largest catalogue of blazar candidates is BROS \citep{itoh2020}, including $88 \, 211$ sources with flat radio spectrum at declination $\delta>-40\degr$ and outside the Galactic Plane ($|b| > 10 \degr$). 
Optical magnitudes are available in $g$, $r$, and $i$ bands 
for nearly half of the sample.
Among the 32980 BROS sources with given $r$-band magnitude, only 12 sources exceed $r=24.5$, the minimum single-visit depth of Rubin-LSST \citep{ivezic2019}.
The positions in the sky of the objects of the BROS and CRATES catalogues are shown in Figure \ref{fig:mappa}.
Table \ref{tab:match} displays the number of sources found by cross-matching the BZCAT5, BROS and CRATES catalogues with the DDFs positions. The total number of candidates in the last column takes into account only non-duplicate sources.
The double EDFS DDF has been split into the single components.

\begin{deluxetable*}{lcccc}
\tablenum{1}
\tablecaption{Number of blazar candidates in each DDF (in decreasing declination order) according to the BZCAT5, BROS and CRATES catalogues. The last column reports the total number of non-duplicated sources.\label{tab:match}}
\tablewidth{0pt}
\tablehead{
\colhead{DDF} & \colhead{BZCAT5} & \colhead{BROS} & \colhead{CRATES} & \colhead{$N_{tot}$}
}
\decimalcolnumbers
\startdata
COSMOS   & 1 & 25 & 1 & 26\\
XMM-LSS  & 0 & 46 & 2 & 46\\
ECDFS    & 0 & 29 & 4 & 31\\
ELAIS    & 0 &  0 & 2 & 2\\
EDFS     & 2 &  0 & 9 & 9\\
EDFS     & 2 &  0 & 3 & 3\\
\hline
Total    &   &    &   & 117\\
\enddata
\end{deluxetable*}

In total, there are 117 blazar candidates in the planned DDFs according to the above three catalogues. 
Most of them are faint objects, down to $r \sim 23.1$; 
only two sources are brighter than 15.5 mag, the median saturation limit in $r$ band.
If we consider that three DDFs are out of the sky area covered by BROS, and the depth that Rubin-LSST can reach, the overall number of expected blazar candidates in DDFs further increases.
The possibility to follow these objects with much more sampling in both light curve and colours than those in the WFD, and with limited saturation issue, makes them a small golden sample to investigate, with important scientific return.

Taking into account the BROS sky coverage, we can estimate an average number density of about 3.30 sources per square degree, which would imply $\sim 136 \, 000$ blazar ``candidates" in the whole sky, about $66 \, 000$ of which in the $\sim 20 \, 000$ square degrees covered by Rubin-LSST, and $\sim 190$ in the six DDFs.
We can wonder whether this number makes sense; in other words, does the blazar population that will be seen by Rubin-LSST reasonably count several tens of thousands objects?
To make a rough estimate, we can consider that the quasar population explored by Rubin-LSST is expected to include about $10^7$ objects up to $z \sim 7$, for a few millions of which multiband light curves will be available to study quasar variability \citep{ivezic2016,ivezic2019}.
A fraction $\zeta$ of these quasars will be radio-loud. If one of their two jets were within $\sim 30\degr$ from the line of sight, we would see a blazar.
Therefore, we can obtain an estimate of the number of blazars that Rubin-LSST will likely observe by scaling the number of radio-loud quasars by the ratio between the solid angle corresponding to a $60 \degr$ aperture and that of the half sphere (which is $2 \pi$): 
\begin{equation}N \approx 10^7 \times \zeta \times 2 \pi (1-\cos(\pi/6)) /(2 \pi)
\end{equation}
The radio-loud fraction $\zeta$ is commonly assumed around $10\%$ \citep[e.g.][]{ivezic2002}, but actually it was found to decrease from about 24\% to roughly 4\% with increasing redshift and decreasing luminosity \citep{jiang2007,kratzer2015}. By adopting the latter value, we would obtain a lower limit of about 53 600 objects, which is of the order of the BROS extrapolation. 
By considering a Rubin-LSST footprint of $\sim 20 \, 000$ square degrees, we would obtain $\sim 2.68$ sources per square degree, and about 150 sources in the six planned DDFs, most of which will presumably be detected with a single-visit exposure.

\section{Saturation}
\label{sec:saturation}
\begin{figure*}
\gridline{\fig{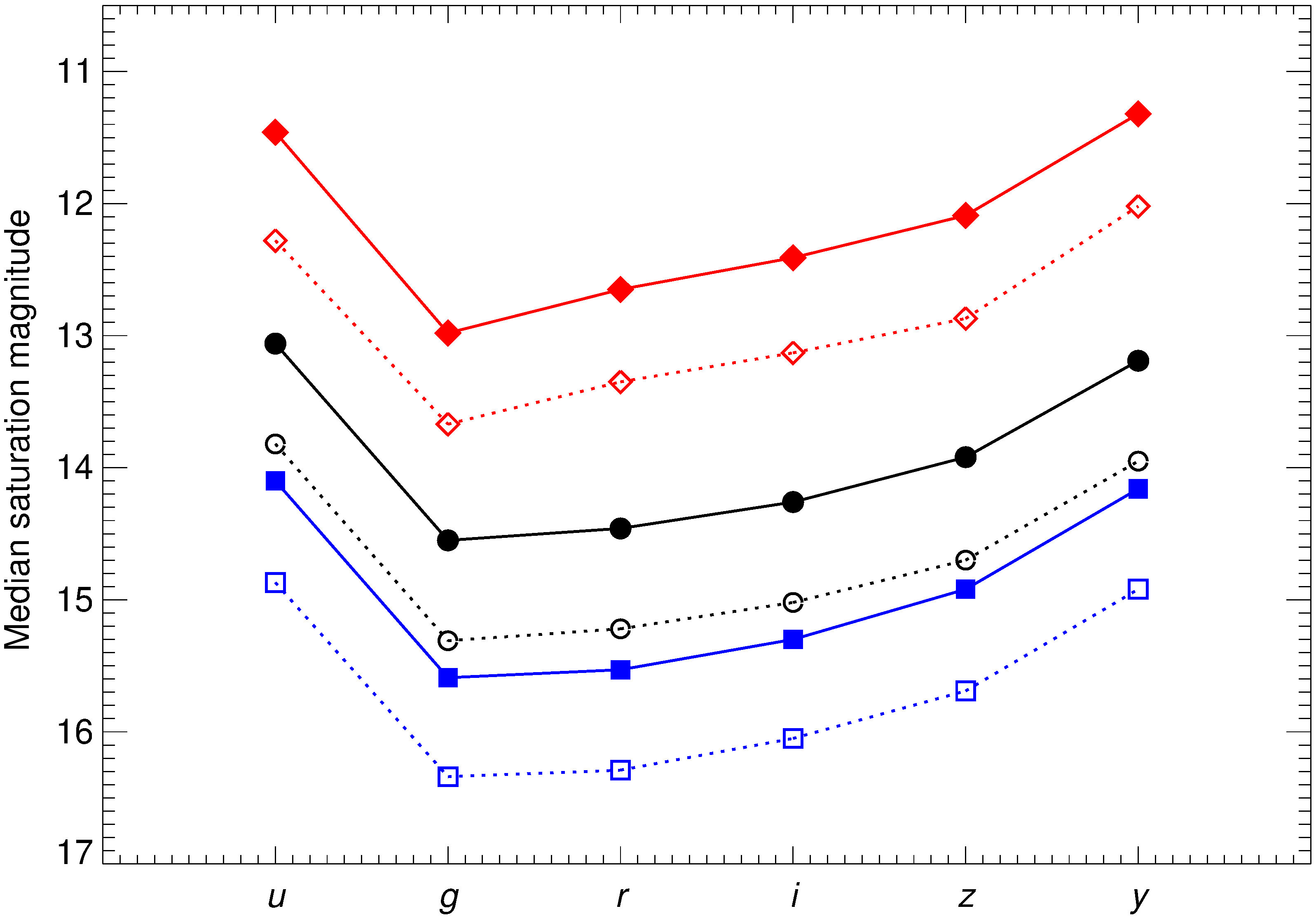}{0.5\textwidth}{(a)}
          \fig{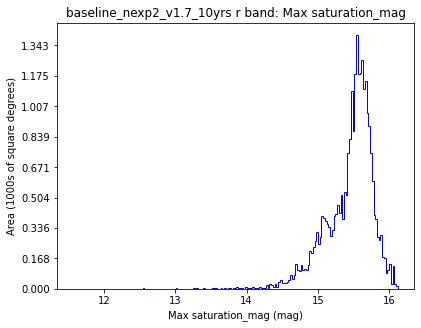}{0.5\textwidth}{(b)}
          }
\gridline{\fig{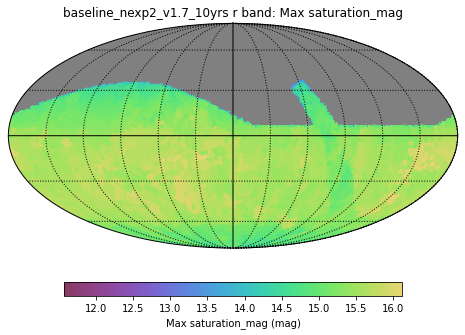}{0.5\textwidth}{(c)}
         }
\caption{(a) Median values over the sky of the maximum (blue squares), median (black dots) and minimum (red diamonds) saturation magnitudes over the 10 yr period of the survey in the different Rubin-LSST filters. Filled symbols refer to the \texttt{OpSim} run \texttt{baseline\textunderscore nexp2\textunderscore v1.7\textunderscore 10yrs}, which includes two exposures of 15 s each per visit; empty symbols to  \texttt{baseline\textunderscore nexp1\textunderscore v1.7\textunderscore 10yrs}, with a single exposure of 30 s per visit.
Histogram (b) and skymap (c) of the maximum saturation magnitude in the $r$ band for the nexp2 cadence simulation.
\label{fig:saturation}}
\end{figure*}

The Rubin-LSST photometry of the brightest blazars, or blazars during outburst states, can be affected by saturation.
In photometric conditions and 0.7 arcsec seeing, the saturation levels for a 15 s exposure are expected to be 14.7, 15.7, 15.8, 15.8, 15.3 and 13.9 mag for the $u, g, r, i, z, y$ bands, respectively\footnote{From the Rubin-LSST Science Book; https://www.lsst.org/scientists/scibook, \citet{lsstsb}}. These values increase with the exposure time.
But since the saturation levels change with observing conditions, they can vary significantly in time and with position in the sky. 
Figure \ref{fig:saturation} shows the median (over the sky) saturation magnitudes obtained with both the \texttt{OpSim} run \texttt{baseline\textunderscore nexp2\textunderscore v1.7\textunderscore 10yrs}, which includes two snapshots with 15 s exposures for all visits, and with the \texttt{baseline\textunderscore nexp1\textunderscore v1.7\textunderscore 10yrs} cadence, where visits consist of single exposures of 30 s. 
For each band, the maximum, median and minimum saturation magnitudes obtained from the simulated 10-yr period of the survey (which includes variable observing conditions) are given.
In the 30-s single exposure case, the level of saturation increases by about 0.7--0.8 mag with respect to the 15-s double exposures.
For a better comprehension of these values, Figure \ref{fig:saturation} also includes the sky map
showing the faintest objects that could saturate in $r$ band, and the corresponding histogram,
for the \texttt{nexp2} cadence simulation.

To investigate the problem of saturation, we can limit ourselves to the BZCAT5 catalogue, where the brightest and more active sources are probably all included, and which offers the advantage of providing a value for the optical brightness.
Although the catalogue magnitude for a variable object represents just one of its possible brightness states, we will use these magnitudes as reference values, with the idea that they cover the whole range of brightness states of the known blazar population.
The 191 sources with $R$-band catalogue magnitude brighter than 15.5, which represents the median maximum saturation level in the Rubin-LSST $r$ band, are highlighted in Figure \ref{fig:mappa}. They would saturate in 15 s exposures, when observed in the best seeing conditions.

\begin{figure*}[ht!]
\gridline{\fig{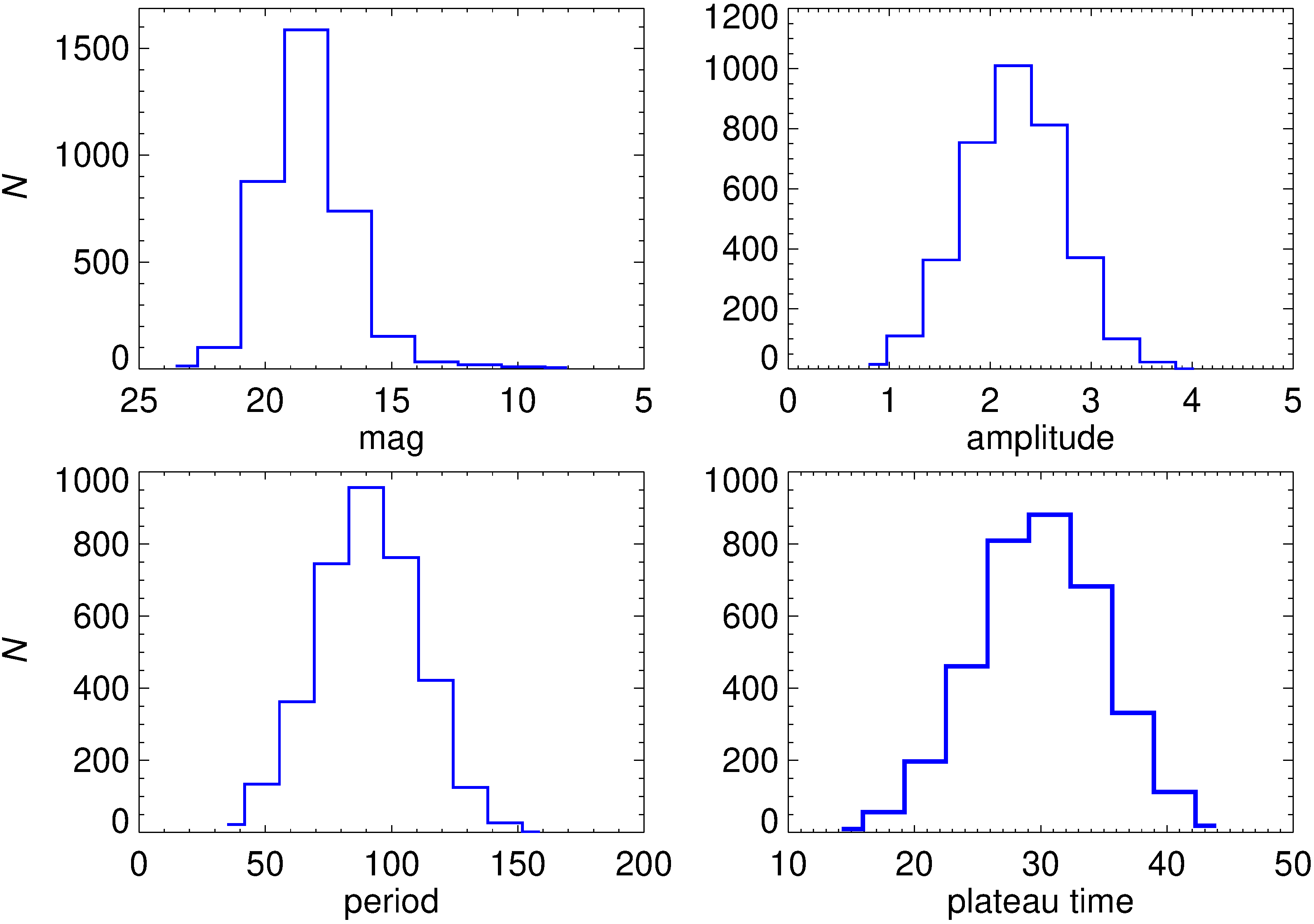}{0.45\textwidth}{}
          \fig{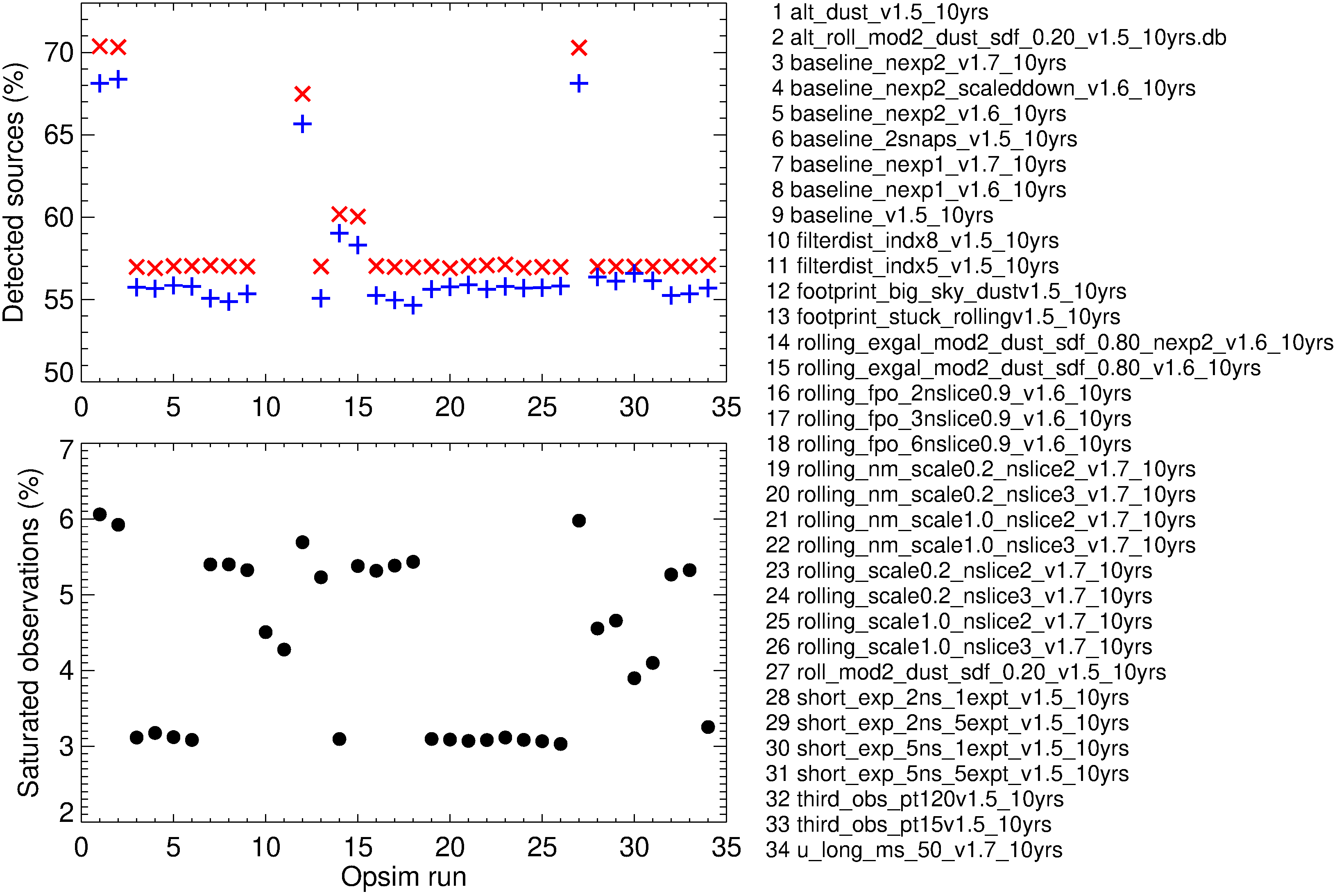}{0.55\textwidth}{}}
\caption{Left: Parameter distributions for the flaring sources used in the saturation metric; from top
left in clockwise direction: the $R$-band catalog magnitudes adopted as ``off-flare" brightness, the outburst amplitude (mag), the plateau time (days), the period (days). 
Right: The top panel shows the percentage of detected sources (red) and of sources detected in outburst (blue), while the bottom panel that of saturated observations, for various \texttt{OpSim} runs listed on the right. \label{fig:simulation}}
\end{figure*}

We performed a simulation (\texttt{BlazarSaturationMetric}) where all the blazars from BZCAT5  are flaring according to given prescriptions.
They go ``on flare"  starting from the catalogue magnitude, which is considered to be the ``off flare" state.
The flare amplitude, temporal distance between flares, and flare duration are randomly chosen from a normal distribution in a range of reasonable values. 
Figure \ref{fig:simulation}
shows the distribution of (i) the blazar catalogue magnitudes used as ``off flare" brightness level; (ii) the flare amplitudes, corresponding to the (maximum - minimum) magnitude; (iii) the time interval in days that the source remains ``on flare" (plateau time); (iv) the time interval in days between consecutive flares (period). 
The same figure also shows the percentage of sources detected (100\% would be reached if Rubin-LSST could see the whole sky) and of sources detected in outburst (i.e.\ that are observed to undergo $\Delta \rm mag > 0.5$ with good photometry).
Finally, it displays the percentage of saturated observations for various \texttt{OpSim} runs (see Section \ref{sec:cadences}).

The largest number of detected sources and sources detected in outburst are obtained in the runs identified as \texttt{alt\textunderscore dust},
\texttt{alt\textunderscore roll\textunderscore mod2\textunderscore dust}, \texttt{footprint\textunderscore big\textunderscore sky\textunderscore dust},  and \texttt{roll\textunderscore mod2\textunderscore dust}. All these simulations imply an extension of the survey footprint.

The number of saturated observations
goes from $\sim 3\%$ to $\sim 6\%$ and the \texttt{OpSim} runs with single exposures of 30 s perform worse than runs with two exposures of 15 s, as expected. Intermediate saturation levels are obtained with the \texttt{short\textunderscore exp} cadences, which include a minisurvey with either 1 or 5-s exposures, either twice or five times per year, and by the \texttt{filterdist} cadences, where the filter distribution is changed with respect to the \texttt{baseline} simulation.

\begin{figure}[ht!]
\plottwo{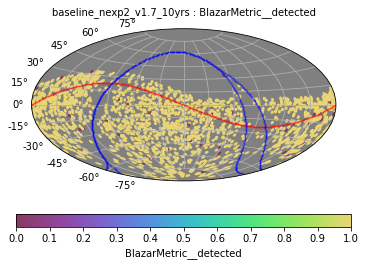}{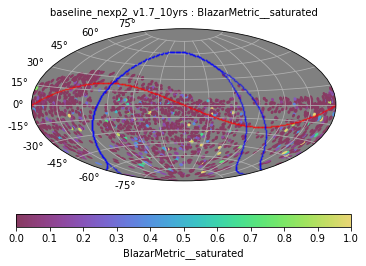}
\caption{Left: Sky map of 2029 simulated blazars. Values of 1 denote they would be
detected brightening by 0.5 mag in the \texttt{baseline} strategy, values of zero mean no detection. 
Right: The fraction of observations per source
that would be saturated.
\label{fig:simsky}}
\end{figure}

In Figure \ref{fig:simsky} we plot the sky maps highlighting whether a source has been observed to flare (left) and
the fraction of saturated observations (right) obtained with the \texttt{baseline\textunderscore nexp2\textunderscore v1.7\textunderscore 10yrs} \texttt{OpSim} run.

We stress that although the saturation
seems to affect only a small fraction of outbursts, these bright events are the most important ones to understand the jet physics, especially in a multiwavelength context. Saturation would hamper the possibility to trigger follow-up observations, and thus to exploit the synergy between Rubin-LSST and high-energy facilities to understand the role played by leptonic versus hadronic processes in producing the X-ray and $\gamma$-ray radiation in blazars, and in turn also the emission of high-energy neutrinos.

\section{Flux and colour sampling}

\subsection{Baseline cadence}
\begin{figure}[ht!]
\plottwo{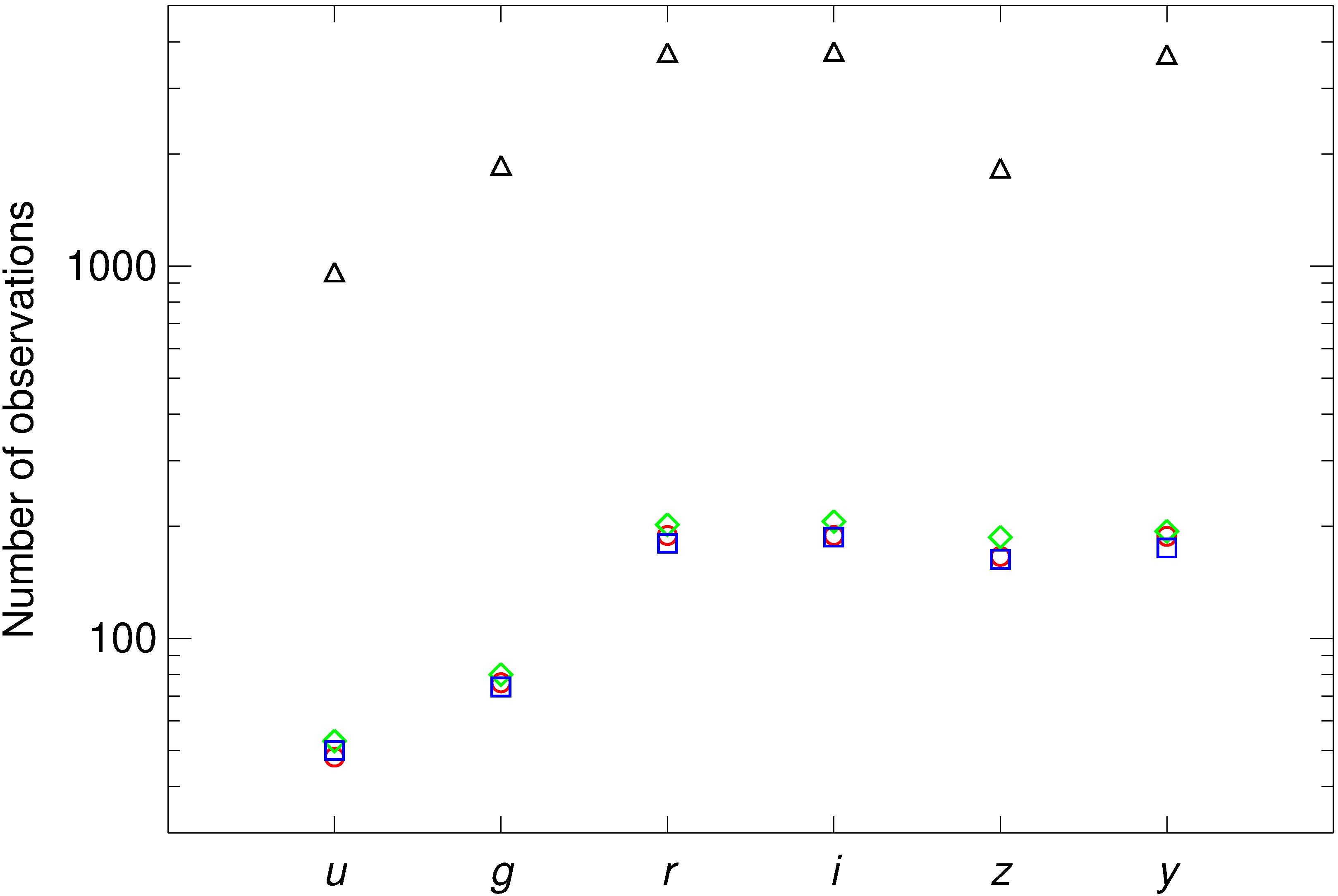}{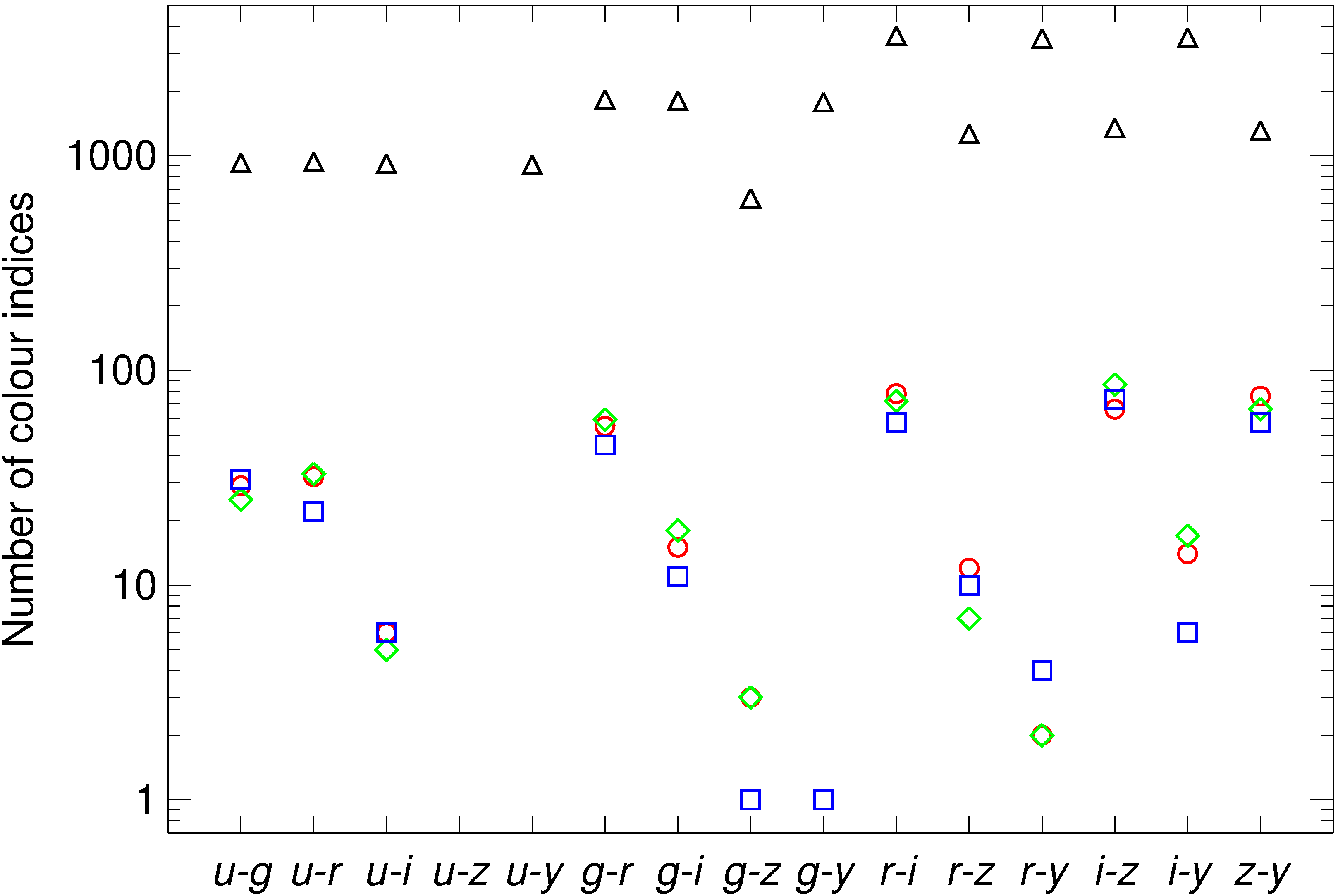}
\caption{The number of observations (left) and the number of colour indices obtained by coupling data in the two bands taken in the same night (right) at the end of the 10-yr survey. 
Values have been obtained using the \texttt{baseline\textunderscore nexp2\textunderscore v1.7\textunderscore 10yrs} \texttt{OpSim} run. Black triangles refer to the DDF, while the other symbols and colours to the WFD in three different positions in the sky.  \label{fig:points}}
\end{figure}

The study of blazar variability requires well-sampled light curves and colour indices.
We first explore the number of data points that can be obtained by both WFD and DDFs with the \texttt{baseline\textunderscore nexp2\textunderscore v1.7\textunderscore 10yrs} cadence simulation in ten years. 
We also investigate the number of colour indices that can be built with data acquired in the same night in order to mitigate a possible bias introduced by variability.
The results are shown in Figure \ref{fig:points} for both the DDF and WFD. In the latter case, we tested three different sky positions. The numbers are brightness-agnostic; they do not depend on the light curve features (provided that all observations lead to a detection) and do not take saturation into account.
The light curve sampling is best in the $r, i, z, y$ filters in the WFD, and in the $r, i, y$ bands in the DDF. The number of points in the DDF can be more than 20 times larger than in WFD. 
The best-sampled colours are: $g-r$, $r-i$, $i-z$, $z-y$ for the WFD, and $r-i$, $r-y$, $i-y$ for the DDF. The maximum number of colour index measurements with data in the same night in the WFD is less than 100 in 10 yr, while in DDFs we have more than 3500 indices in the best cases. 

\begin{figure}[ht!]
\plottwo{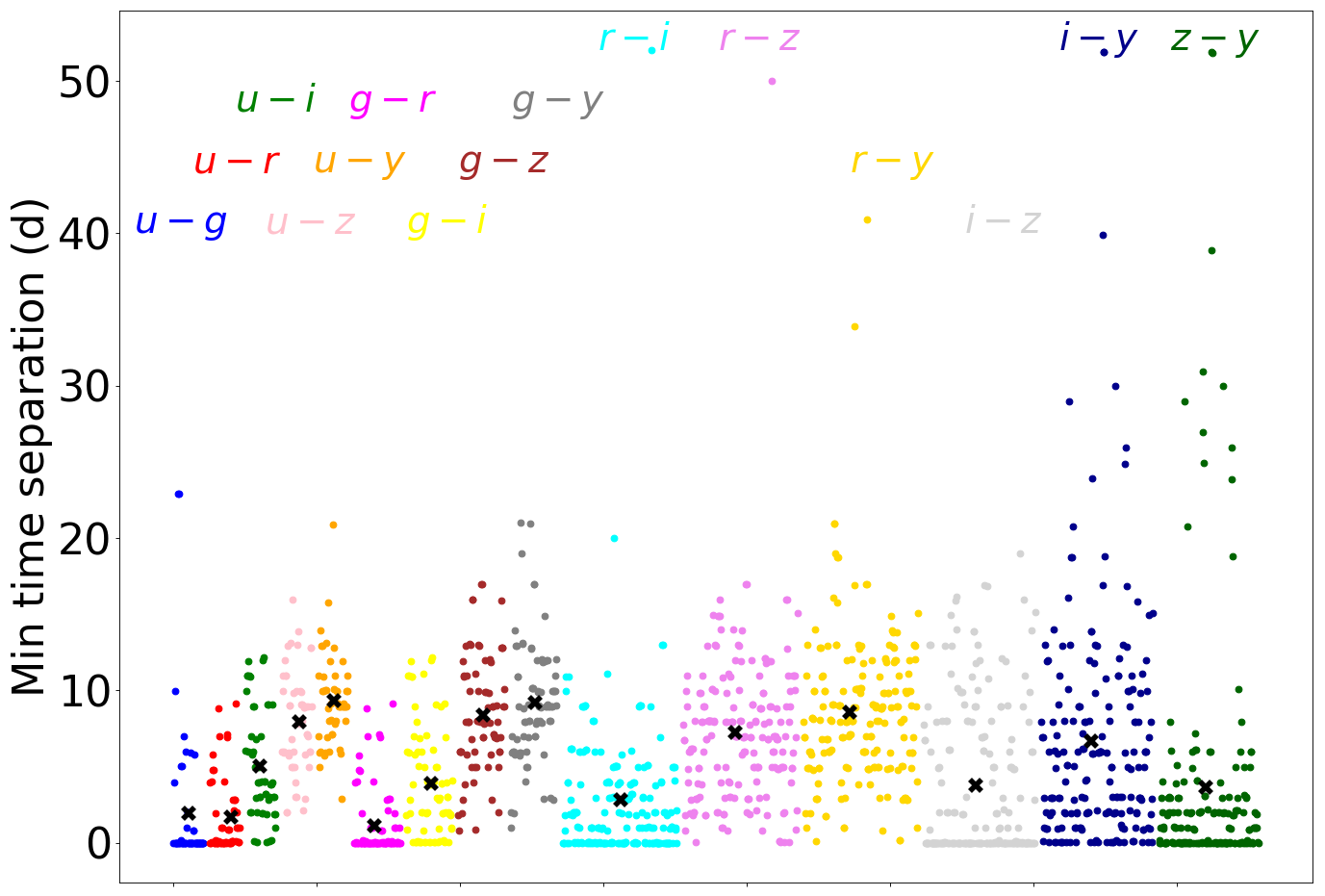}{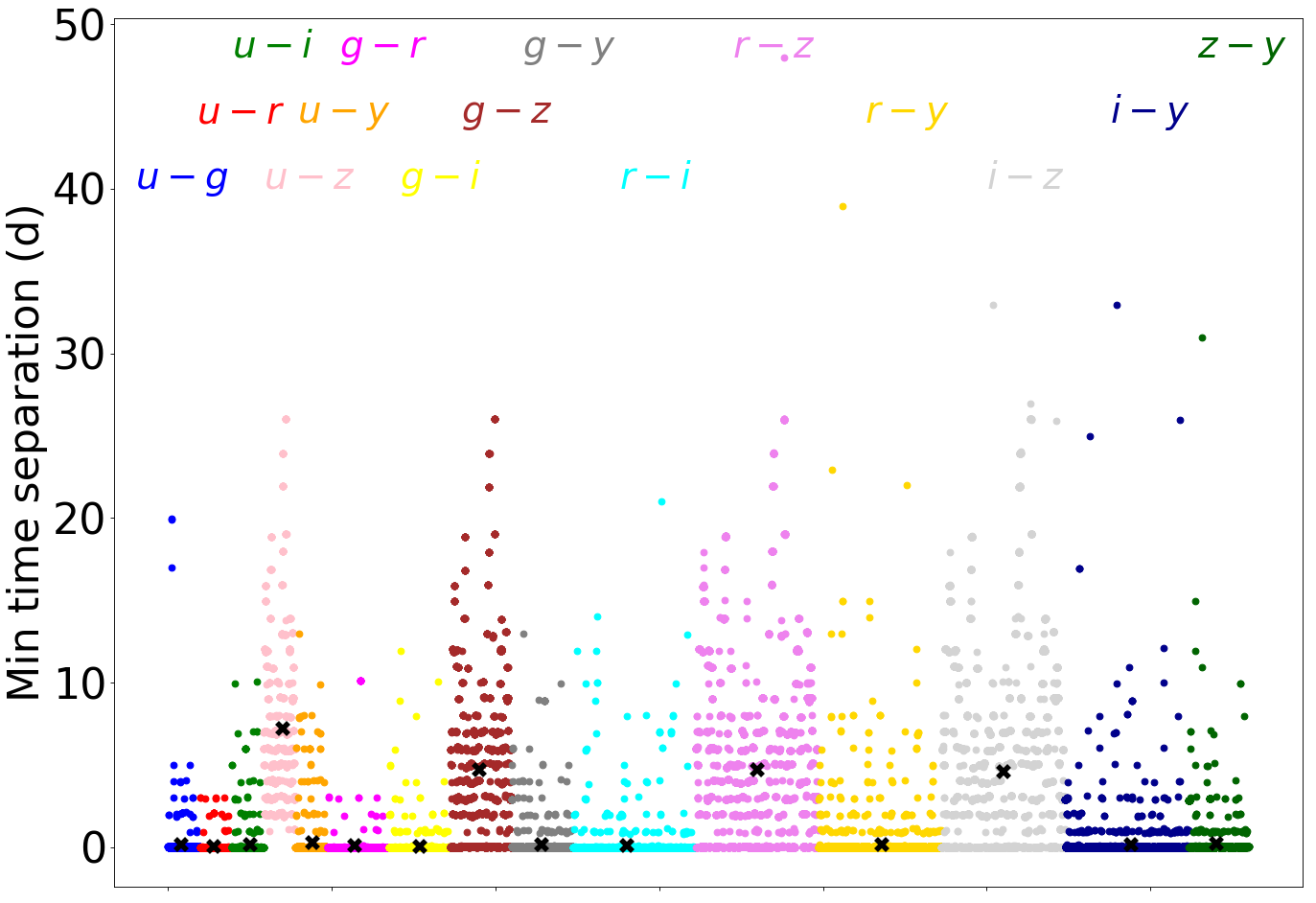}
\caption{The minimum time separation in days between observations in two different filters for the WFD (left) and the DDFs (right).
Values have been obtained using the \texttt{baseline\textunderscore nexp2\textunderscore v1.7\textunderscore 10yrs} \texttt{OpSim} run. Different colours are used for different filter pairs, and the horizontal width of each pair is proportional to the corresponding number of points to better distinguish the number density. Black crosses indicate mean values. \label{fig:minti}}
\end{figure}

We can wonder how much we should enlarge the time separation between observation in two different filters to either significantly improve the number of corresponding colour index measurements and thus trace the spectral changes more efficiently, or even to obtain colours that cannot be obtained with the same-day constraint.
Figure \ref{fig:minti} displays the distribution of the minimum time separation between the acquisition of data in two different bands.
Results for both the WFD and the DDFs are shown.
The choice of the time separation limit should depend on how fast the variability of the source is: while for OVV sources the same-day constraint seems a necessary requirement, for the other (more slowly variable) objects a few-day time separation may be acceptable.

For the WFD, the \texttt{baseline} cadence implies a median value (over the sky) of 3.83 days for the median (over 10 yr) inter-night time gaps between consecutive observations of the same field, when all bands are considered. 
When the single bands are distinguished, the median time gaps are 21.94, 18.46, 7.87, 9.02, 11.89, and 20.85 days in the $u, g, r, i, z$ and $y$ bands, respectively.  

\begin{figure}[ht!]
\plottwo{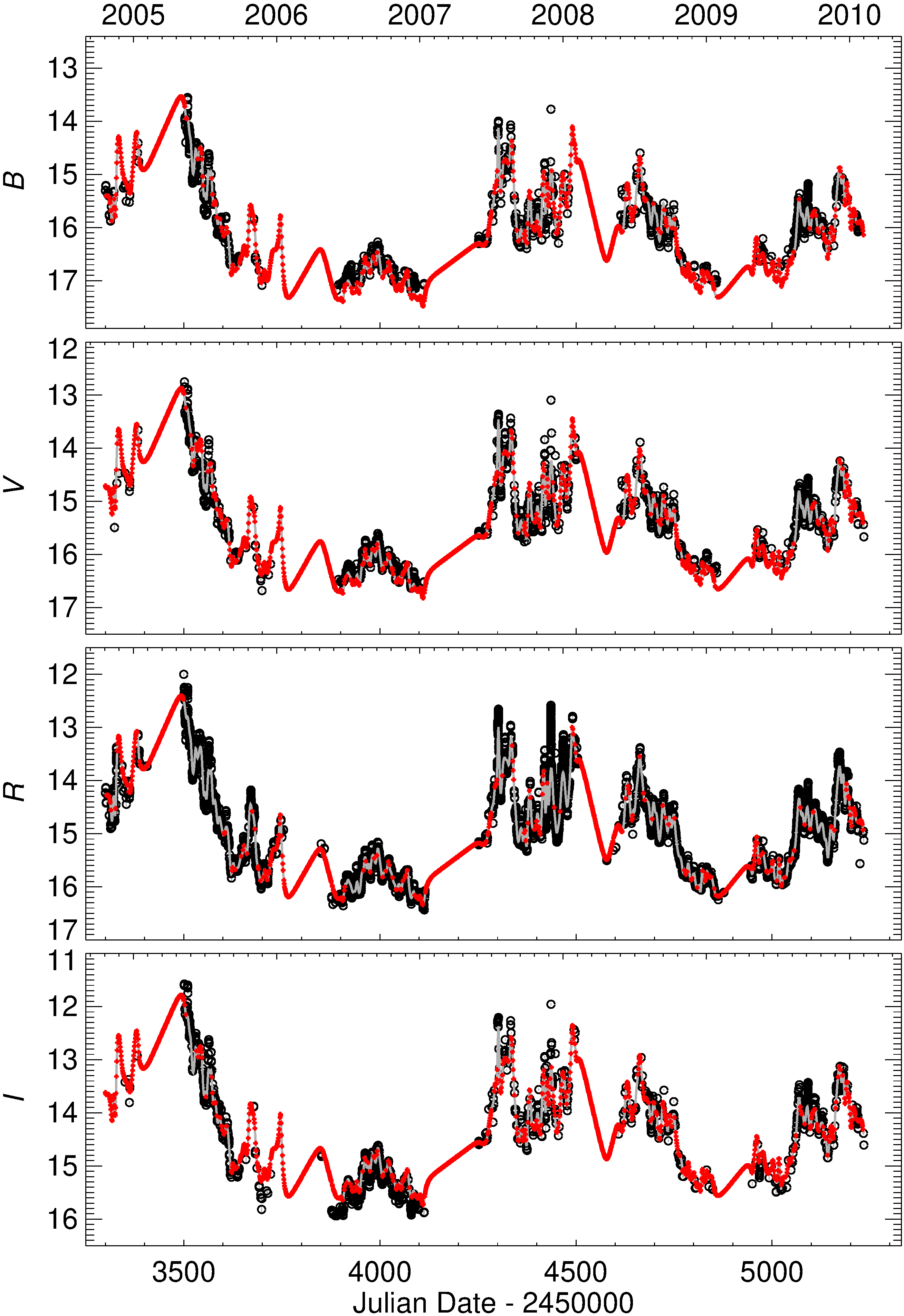}{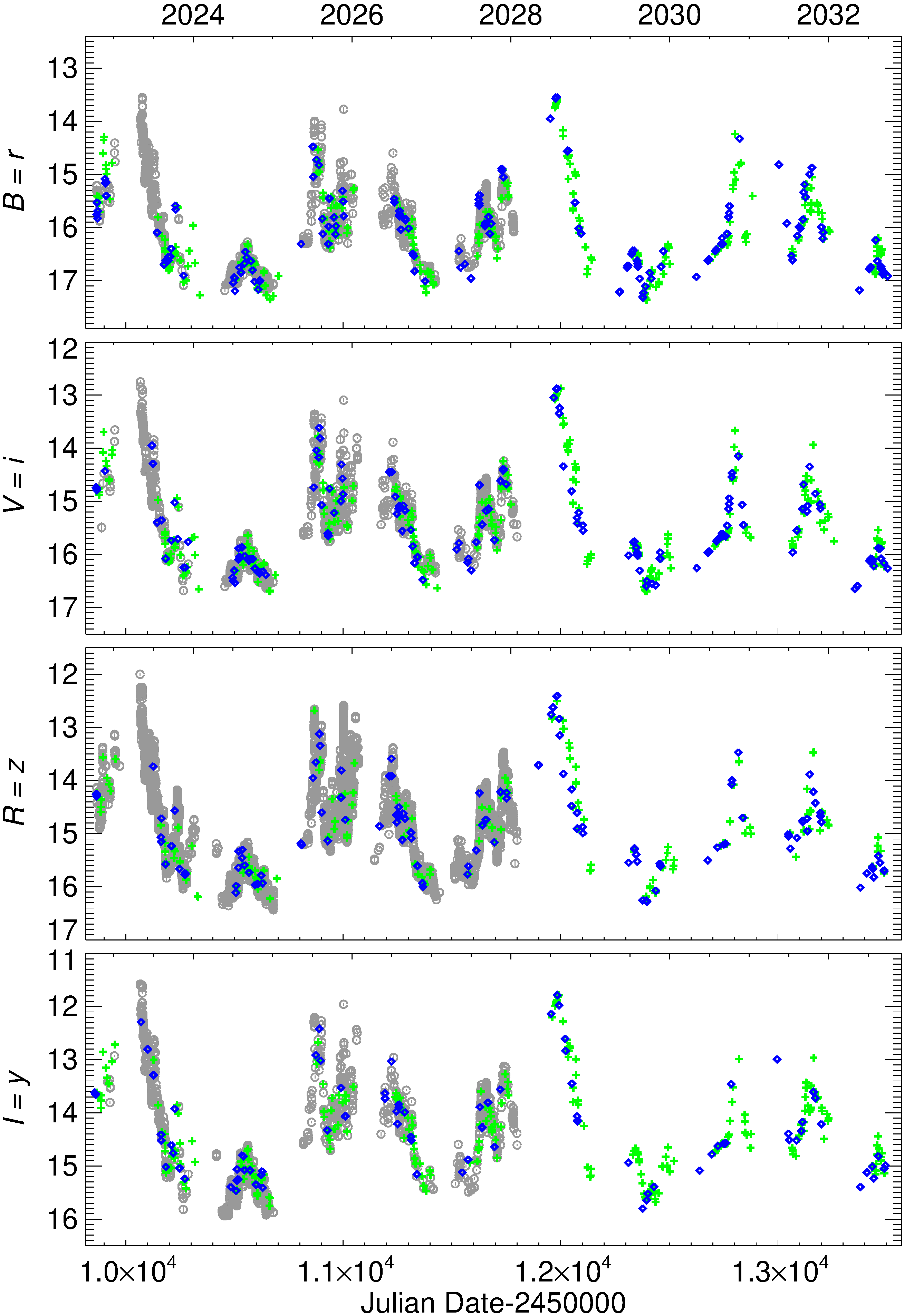}
\caption{Left. The $BVRI$ light curves of the blazar 3C 454.3 processed by the WEBT. The red line in the $R$-band panel represents a cubic spline interpolation through the 4-d binned light curve. The red lines in the other panels are the same cubic spline interpolation shifted according to average colour indices.
Right. The simulated Rubin-LSST resulting sampling according to the \texttt{TransitAsciiMetric} with a 10 yr \texttt{baseline} cadence.  Blue diamonds refer to the WFD, green plus signs to the DDF. Grey symbols indicate the WEBT data already shown in the left panel.\label{fig:crazy}}
\end{figure}

To see the effects of Rubin-LSST sampling in a real case, we used the processed $BVRI$ light curves of the OVV 3C~454.3, which were obtained by the monitoring effort of the WEBT and published in \citet{villata2006,villata2007,villata2009,raiteri2007,raiteri2008,raiteri2008l,raiteri2011}. 
These are shown in Figure \ref{fig:crazy} and cover a time interval of 1980 d, i.e.\ $\sim 5.425 \,\rm yr$.
We adopted the \texttt{baseline\textunderscore nexp2\textunderscore v1.7\textunderscore 10yrs} cadence and the \texttt{TransitAsciiMetric} available in the \texttt{MAF}. 
Since the code makes interpolations between data points, and the sampling of the WEBT light curves is not the same in all bands, we filled the gaps in the light curves in a controlled way in order to avoid different interpolated trends in different bands.
To do this, we first ran a cubic spline interpolation on the 4-d binned $R$-band light curve, which is the best-sampled one. 
Then we calculated the average colour indices by coupling data within 30 min. 
We got: $<B-R>=1.13$, $<V-R>=0.47$, $<R-I>=0.62$. 
These values were used to shift the spline to match the light curves in the other bands. 
We gave as input to the \texttt{TransitAsciiMetric} the $BVRI$ light curves containing WEBT data with errors less than 0.1 mag, complemented by spline data (shifted according to the corresponding mean colour index) in the days where real data are not available (see Figure \ref{fig:crazy}).
Since the Rubin-LSST most sampled light curves are in the $r,i,z,y$ bands, we established the correspondence $B=r$, $V=i$, $R=z$, and $I=y$.
Because the cadence simulation is run for the whole survey length, at the end of the period covered by the WEBT observations, the light curves are repeated again from the beginning by the code to cover the entire 10-yr period. 

The simulated light curves obtained by running the \texttt{baseline OpSim} on the 3C 454.3 interpolated WEBT light curves are plotted in Figure \ref{fig:crazy}, where both the WFD and DDF sampling are considered.
We note that the counterpart of the peak of the outstanding outburst of 2005 in the 10-yr simulation would go undetected by Rubin-LSST in 2023, but would be fairly sampled in 2028, even by the WFD.

\begin{figure}[ht!]
\plottwo{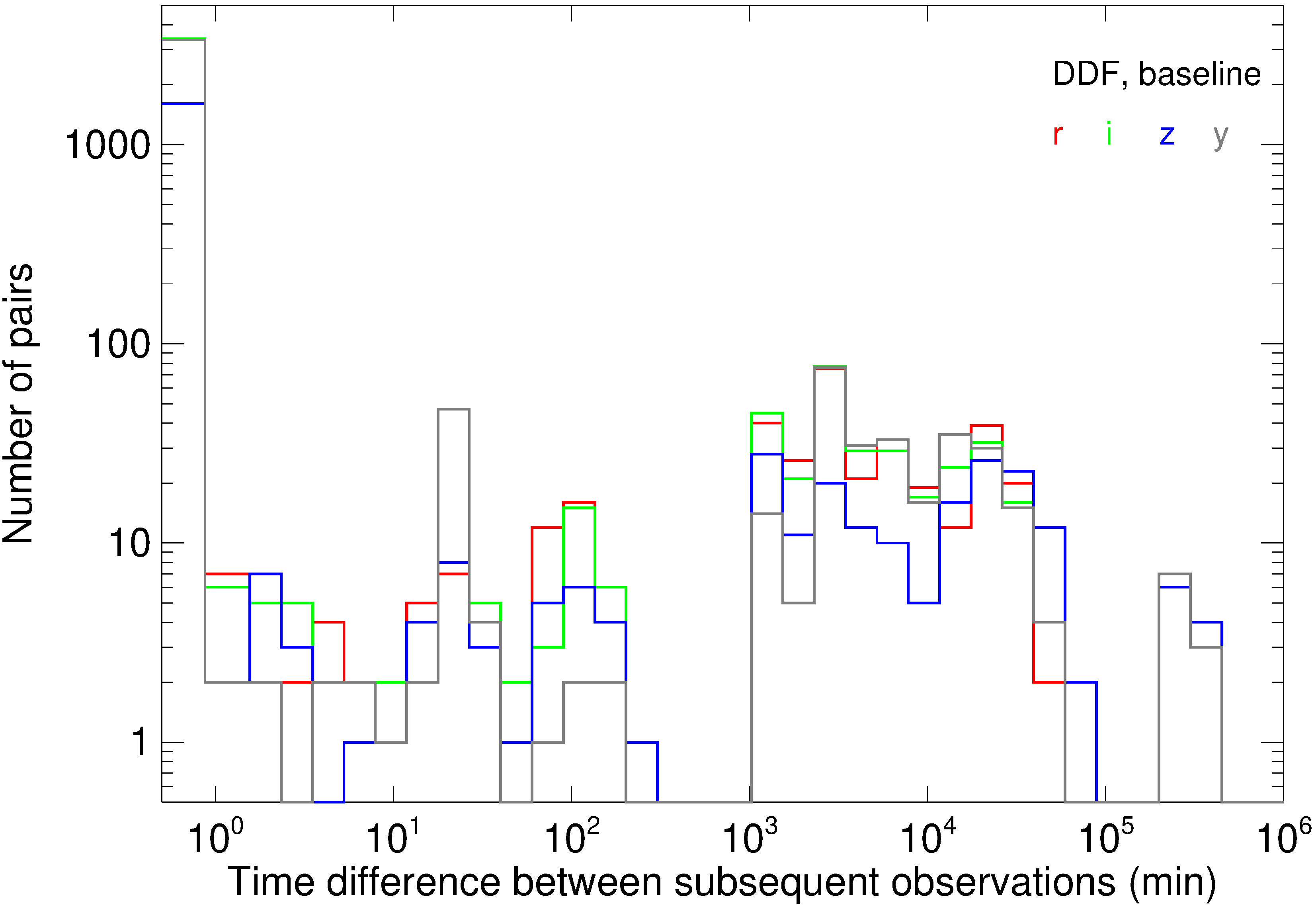}{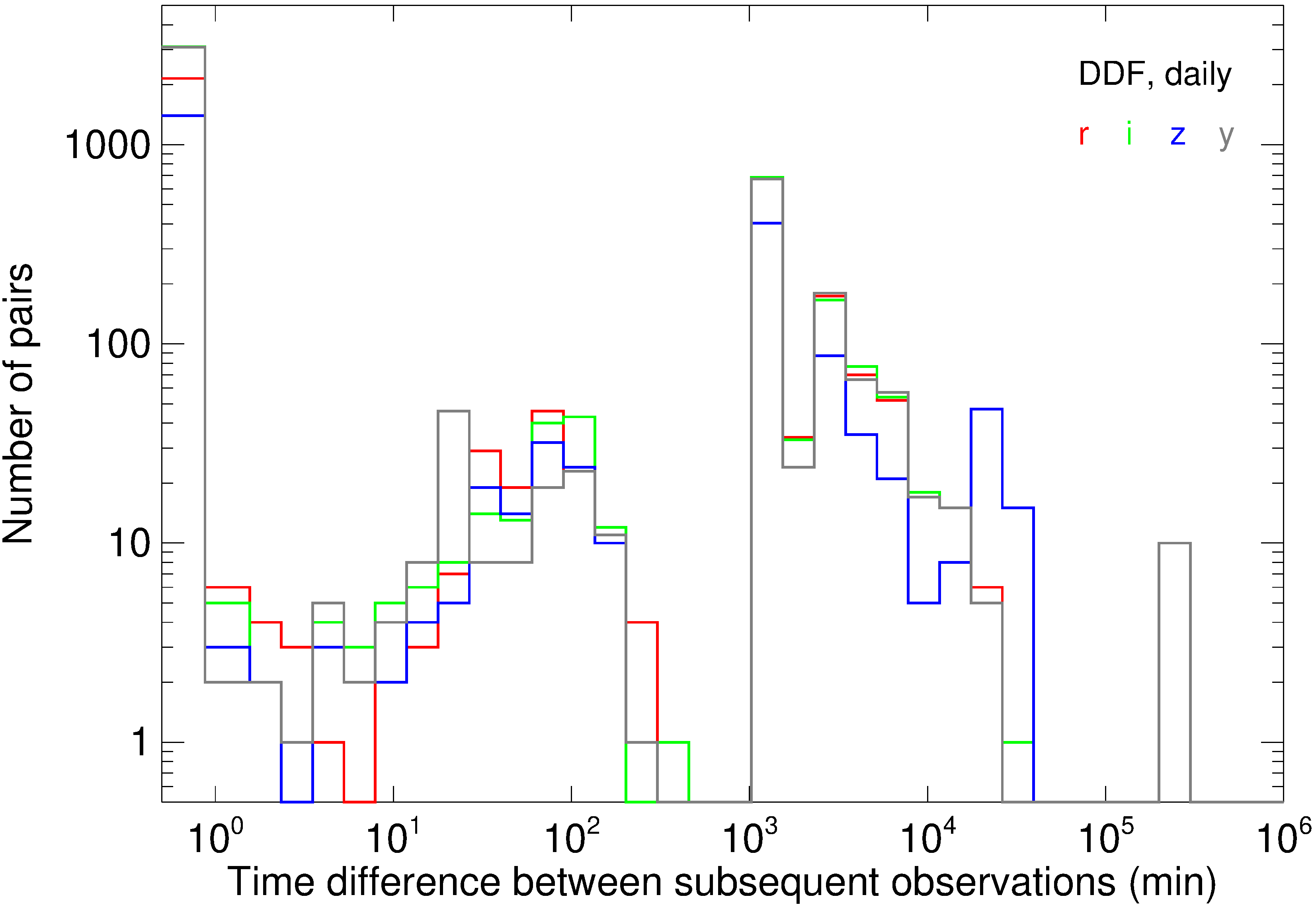}
\caption{Time separation (min) between subsequent observations of a DDF in the same filter. 
Different colours distinguish different bands, as indicated in the legend. The left panel refer to the \texttt{baseline\textunderscore nexp2\textunderscore v1.7\textunderscore 10yrs} cadence simulation and the right panel to the \texttt{daily\textunderscore ddf\textunderscore v1.5\textunderscore 10yrs} one.\label{fig:dt}}
\end{figure}

Table \ref{tab:lcsamp} reports the number of data points in the simulated $rizy$ light curves of Figure \ref{fig:crazy}, and the corresponding number of observing nights, to highlight the intra-night sampling. 
In the WFD, the average number of points per night is 1.3-1.4 in the $r,i,z$ bands, and it is 2.1 in $y$, implying more intra-night observations in this filter. These values rise to 10--14 data points per night in the DDF. 
Therefore, if the number of data points in the DDF is 11-20 times larger than that in the WFD, the observing nights are only 1.4-3 times more numerous. This is why we cannot see a much better sampling in the DDF with respect to that in the WFD in Figure \ref{fig:crazy}.
Moreover, the total duration of the intranight observing sequence in the same filter is about 11 min, which in general is too short to follow even the fastest flux changes of OVV blazars.
The distribution of the time separation between subsequent DDF observations is shown in Figure \ref{fig:dt}: about 90\% of cases lie in the first bin, indicating a gap of less than 1 min.
Table \ref{tab:lcsamp} and Figure \ref{fig:dt} also show a different choice for the DDF cadence, which will be discussed in Section \ref{sec:others}.

\begin{deluxetable*}{ccccccc}
\tablenum{2}
\tablecaption{Light curve sampling in terms of number of data points and observing nights in the case of the \texttt{baseline\textunderscore nexp2\textunderscore v1.7\textunderscore 10yrs} cadence simulation and for both the WFD and the DDF. For the DDF, we also show the results of the \texttt{daily\textunderscore ddf\textunderscore v1.5\textunderscore 10yrs} cadence.\label{tab:lcsamp}}
\tablewidth{0pt}
\tablehead{
 & \multicolumn2c{WFD (\texttt{baseline})} & \multicolumn2c{DDF (\texttt{baseline})} & \multicolumn2c{DDF (\texttt{daily})}\\
\colhead{Band} & 
\colhead{$N_{\rm points}$} & \colhead{$N_{\rm nights}$} & 
\colhead{$N_{\rm points}$} & \colhead{$N_{\rm nights}$} & 
\colhead{$N_{\rm points}$} & \colhead{$N_{\rm nights}$}
}
\decimalcolnumbers
\startdata
$r$ & 195 & 147 & 3740 & 298 & 3368 & 1058\\
$i$ & 191 & 149 & 3765 & 305 & 4323 & 1065\\
$z$ & 171 & 122 & 1831 & 176 & 2173 & 656\\
$y$ & 184 &  88 & 3699 & 270 & 4274 & 1048\\
\enddata
\end{deluxetable*}

\begin{figure}[ht!]
\plottwo{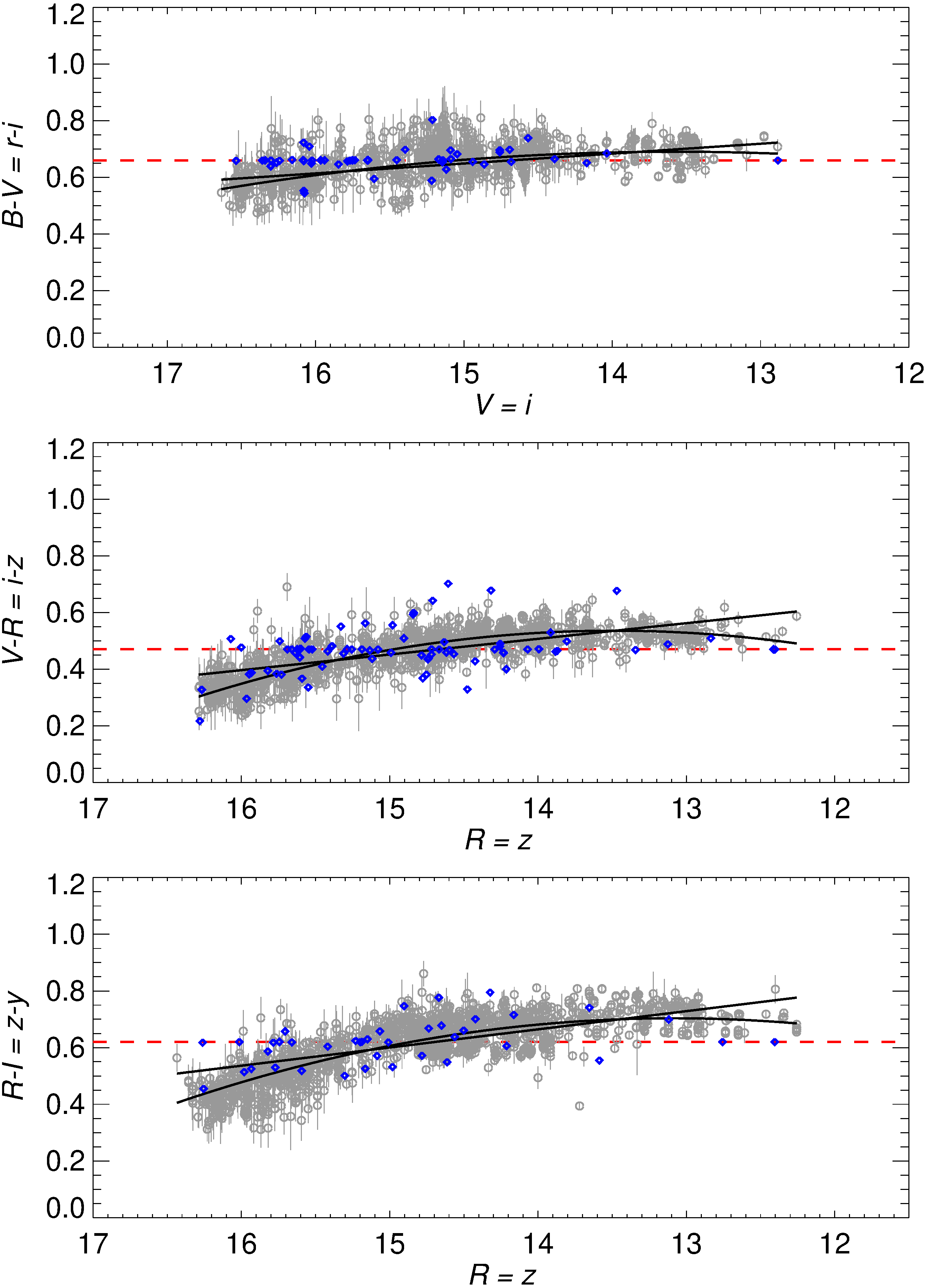}{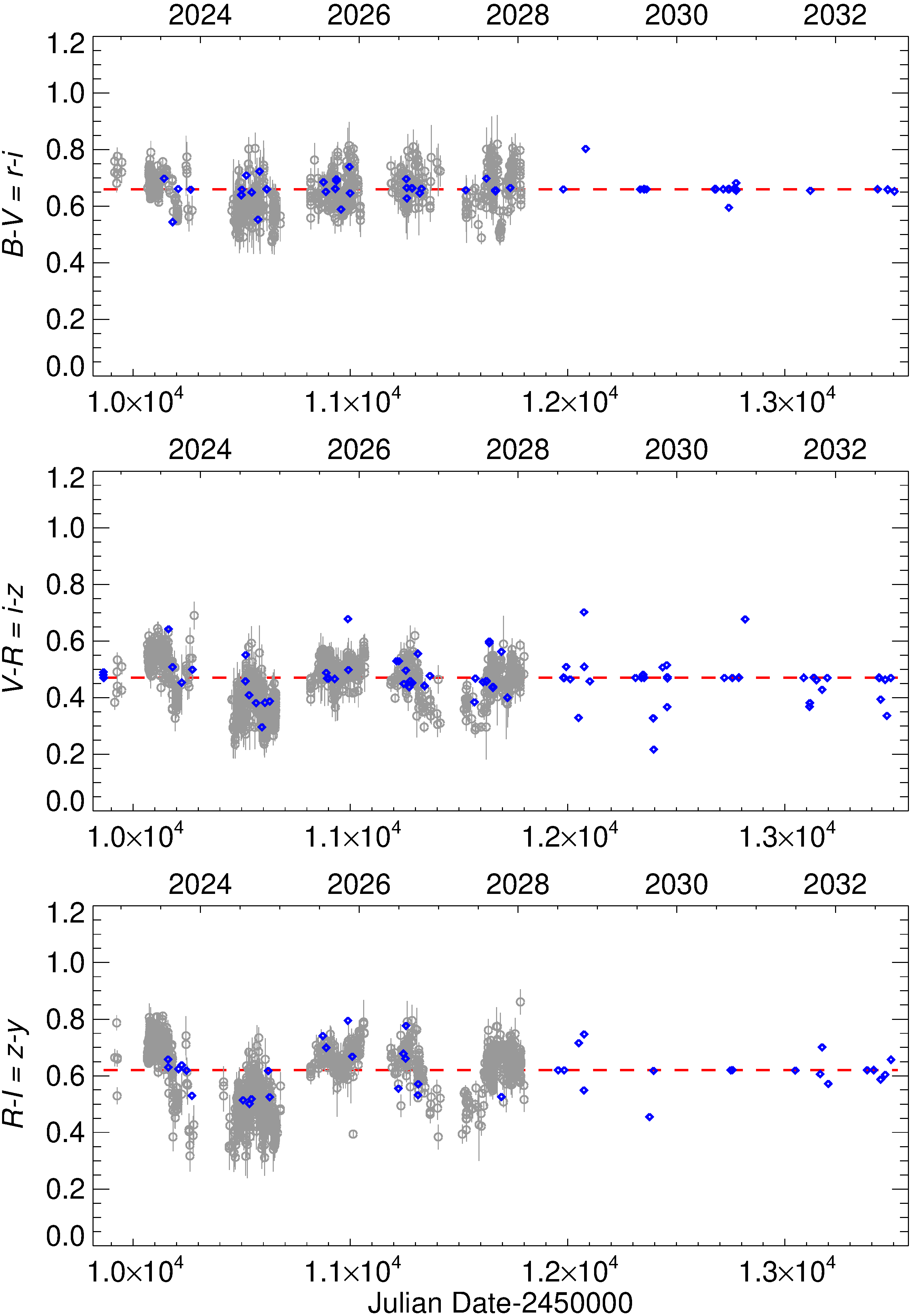}
\caption{Colour indices as a function of brightness (left) and time (right). Grey circles and blue diamonds represent real WEBT and simulated WFD colours, respectively. All of them have been obtained by coupling data with time separation less than 1 hr. The red dashed line indicates the mean colour index derived from WEBT data.
The accumulation of blue points close to the average level is explained in the text.
The black lines are linear and parabolic fits to the WEBT data, to highlight the redder-when-brighter trend with the ``saturation" effect at the bright end. 
All plots have the same amplitude $\Delta \, \rm index=1.2 \, mag$ to make the comparison among colours easier.
A few outliers, deviating more than 3 sigma from the mean, have been discarded. \label{fig:colori}}
\end{figure}

As shown in Figure \ref{fig:points}, the colour indices with better sampling in the WFD are  $r-i$, $i-z$, and $z-y$.
This means $B-V$, $V-R$ and $R-I$ for the WEBT data, according to the correspondence we set.
Figure \ref{fig:colori} shows the colour indices obtained from the WEBT data with time difference between the two filters of 1 h.
As mentioned before, a small time interval is needed because of the strong (and chromatic) intra-night variability of the OVV sources like 3C~454.3. 
In the considered period of 5.4 yr, utilizing the WEBT observations we calculated 1147 $B-V$, 1672 $V-R$, and 1699 $R-I$ indices, with a mean time separation between observations in the two bands of about 7, 10, and 9 min, respectively.
These are compared with the colour indices obtained with the \texttt{baseline} cadence simulation on the interpolated WEBT light curves and the same time separation of 1 h after 10 yr. 
For the WFD, we got 72 $r-i$, 99 $i-z$, and 47 $z-y$ indices, with mean time separations between data in the two bands of 27, 24, and 26 min, respectively.
If we relaxed the time separation to 1 d, the number of indices would become 82, 112, and 73, respectively.
The WEBT and simulated WFD colour indices are shown in Figure \ref{fig:colori} as a function of both brightness and time. The same $\Delta \, \rm index=1.2 \, mag$ in all panels makes a comparison among them easier.

When the two data points yielding the simulated colour index are both taken from the spline interpolation, the resulting colour index is close to the average value used to shift the cubic spline interpolation through the $R$-band data points to match the data in the other bands. This explains the accumulation of data points about the average colour index. 
In any case, it seems that it is hard to sample the spectral changes in the brightest phases and several fast colour variations are lost.
However, after a few years the WFD would be able to assess the main features of blazar spectral variability: the fact that fast variations appear more chromatic than the long-term ones and that there is a trend with brightness. In the case of 3C~454.3, the trend is ``redder-when-brighter" above a certain magnitude ($R \ga 14$) because of the presence of nuclear emission, which is steadier and bluer than the more variable synchrotron radiation. For increasing brightness, a kind of ``saturation" effect appears, as discussed in \citet{villata2006}.
In the figure, linear and parabolic fits have been drawn to highlight this behaviour, which can be better appreciated when the colour involves redder filters, including more variable synchrotron contribution.

\begin{figure}[ht!]
\plottwo{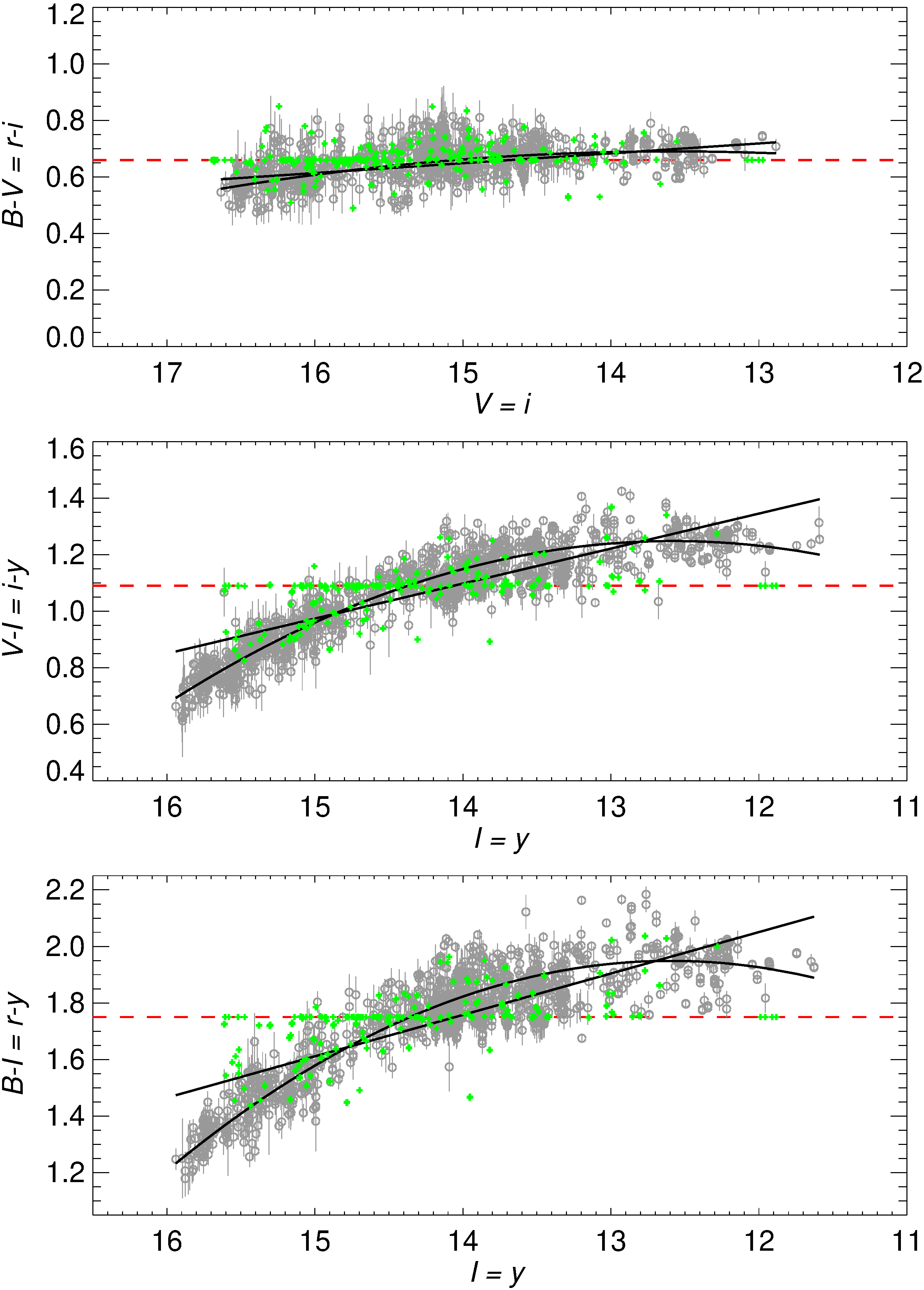}{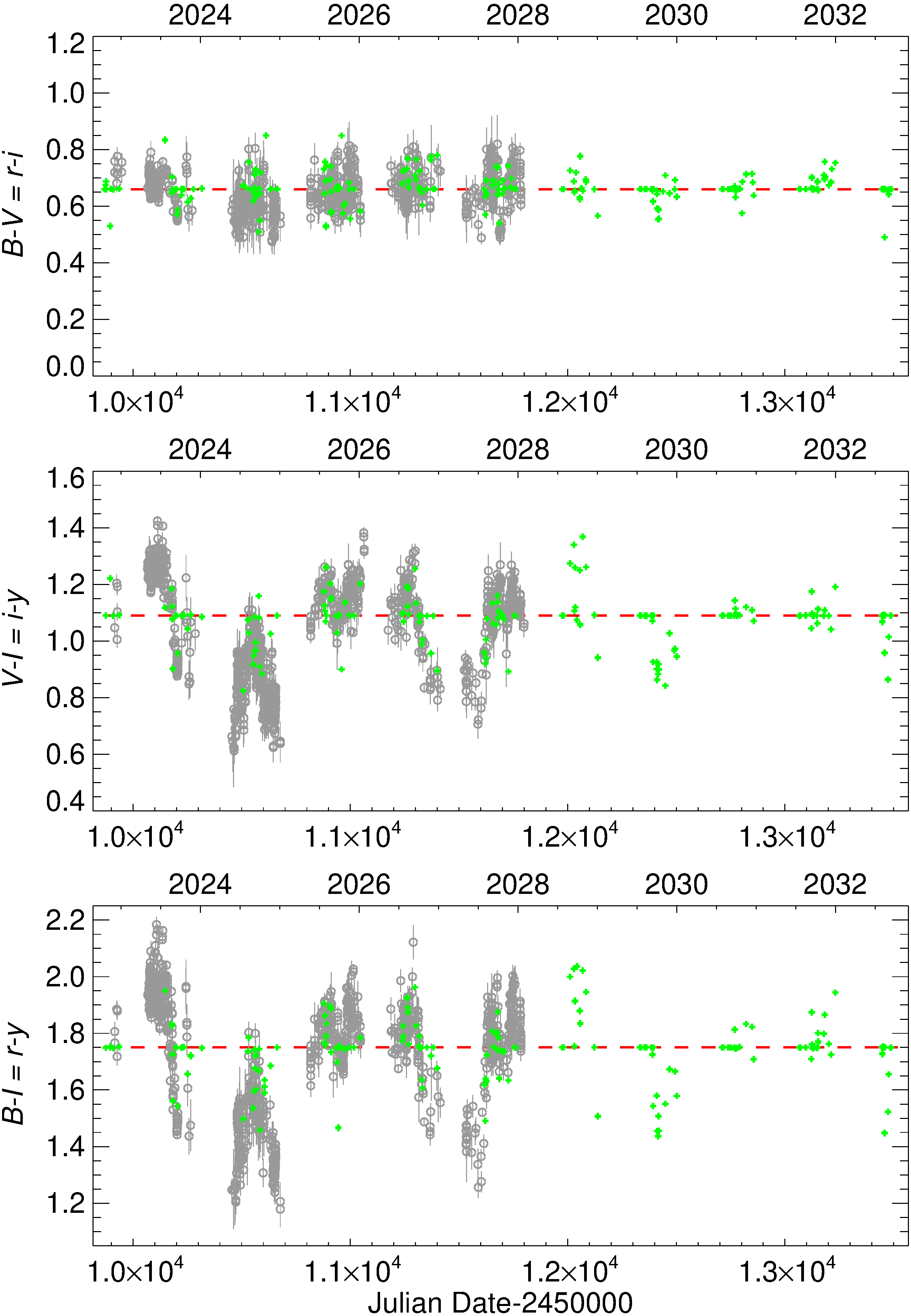}
\caption{Colour indices as a function of brightness (left) and time (right). Grey circles and green plus signs represent real WEBT and simulated DDF colours, respectively. All of them have been obtained by coupling data with time separation below 1 hr. The red dashed line indicates the mean colour index derived from WEBT data.
The accumulation of green points close to the average level is explained in the text.
The black lines are linear and parabolic fits to the WEBT data, to highlight the redder-when-brighter trend with the ``saturation" effect at the bright end. 
All plots have the same amplitude $\Delta \, \rm index=1.2 \, mag$ to make the comparison among colours easier.
A few outliers, deviating more than 3 sigma from the mean have been discarded. \label{fig:colori_ddf}}
\end{figure}

In the DDF \texttt{baseline} cadence simulation, the best colour sampling is reached for $r-i$, $r-y$ and $i-y$, which corresponds to $B-V$, $B-I$, and $V-I$. 
As can be seen from Table \ref{tab:colsamp}, the number of $r-i$, $r-y$, and $i-y$ colour indices that can be obtained in a DDF is 50, 36, and 75 times larger than in the WFD, respectively.
However, the number of nights where they are obtained are only 4.4, 2.3, and 4.5 times larger. As in the case of flux, also for colours the duration of the intranight sampling is too short to follow the spectral variability even of OVV sources.
The $r-i$, $r-y$, and $i-y$ colour indices obtained with the WEBT data by coupling observations within 1 h are displayed in Figure \ref{fig:colori_ddf}, where they are compared to those obtained with the \texttt{baseline} DDF cadence simulation. The same $\Delta \, \rm index=1.2 \, mag$ of Figure \ref{fig:colori} has also been adopted here for an easier comparison of all the colour behaviours.
We note that both the amplitude of the colour variability and the colour trends with brightness are more noticeable when the two bands are more separated.
Indeed, while the $r-i$ colour indices cover only a small range of values and the trend with brightness is weak, the $i-y$ colours show a larger range and a more defined trend, and the $r-y$ colours span almost the whole $\Delta \, \rm index$ range, with linear and parabolic fits of the colour-versus-brightness plot indicating a larger slope and a more pronounced curvature than in the other cases.

As mentioned in the Introduction, the colour variability shown in Figures \ref{fig:colori} and \ref{fig:colori_ddf} is a property of the source and its interpretation can shed light on the emission contributions and physical conditions in the jet. Therefore, it would be present also in case of simultaneous photometry in the two bands, but may be altered by flux variability, if observations in the two bands are too separated in time.

\begin{deluxetable*}{cccccc}
\tablenum{3}
\tablecaption{Colour indices sampling in terms of number of data points and observing nights in the case of the \texttt{baseline\textunderscore nexp2\textunderscore v1.7\textunderscore 10yrs} cadence simulation and for both the WFD and the DDF\label{tab:colsamp}}
\tablewidth{0pt}
\tablehead{
\multicolumn3c{WFD} & \multicolumn3c{DDF}\\
\colhead{Index} & \colhead{$N_{\rm points}$} & \colhead{$N_{\rm nights}$} & 
\colhead{Index} & \colhead{$N_{\rm points}$} & \colhead{$N_{\rm nights}$}
}
\decimalcolnumbers
\startdata
$r-i$ & 72 & 51 & $r-i$ & 3634 & 225\\
$i-z$ & 99 & 76 & $r-y$ & 3531 & 177\\
$z-y$ & 47 & 40 & $i-y$ & 3534 & 180\\
\enddata
\end{deluxetable*}

\subsection{Other cadences}\label{sec:others}
\begin{figure}[ht!]
\plottwo{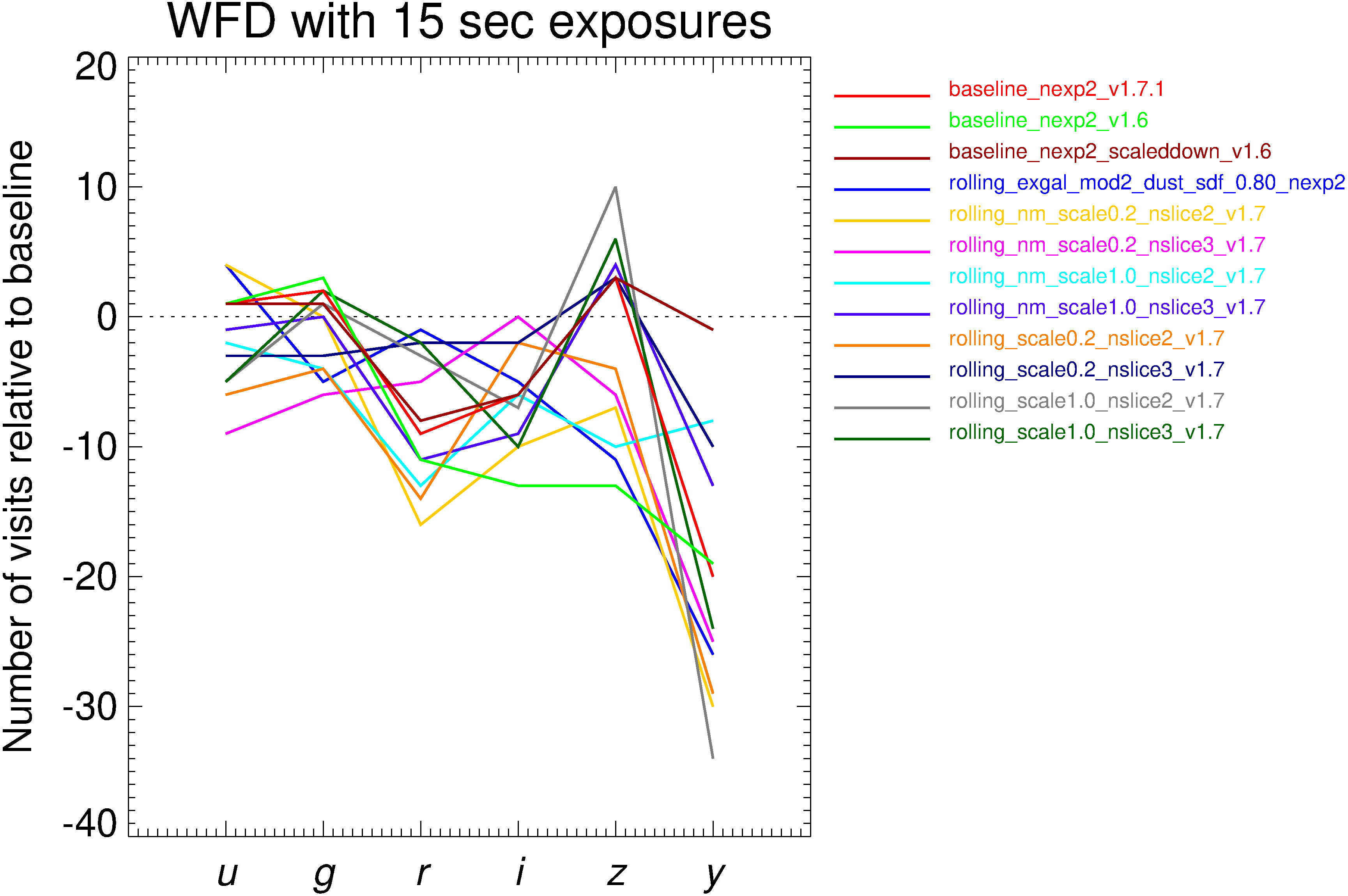}{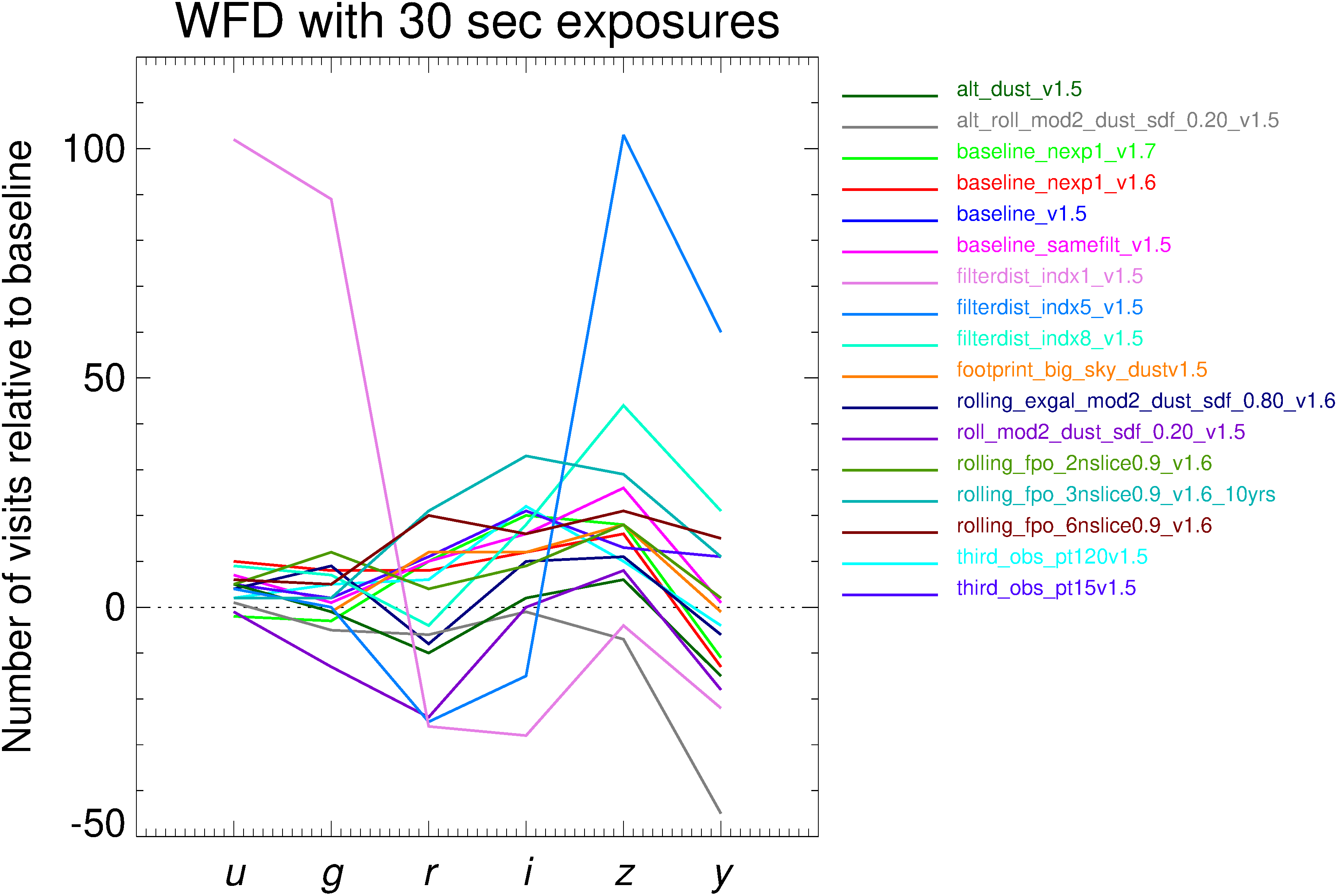}
\caption{Difference between the number of WFD visits in various cadence simulations and that of the \texttt{baseline\textunderscore nexp2\textunderscore v1.7\textunderscore 10yrs} cadence. \texttt{OpSim} runs with either double 15 s (left) or single 30 s (right) exposure times per visit are shown.  \label{fig:filters_wfd}}
\end{figure}

Up to now we have explored light curve and colour sampling according to the \texttt{baseline\textunderscore nexp2\textunderscore v1.7\textunderscore 10yrs} cadence. 
In this section, we investigate whether other \texttt{OpSim} runs can lead to better results.
Figure \ref{fig:filters_wfd} reports the difference between the number of observations obtained over ten years with various WFD cadence simulations, with respect to the \texttt{baseline\textunderscore nexp2\textunderscore v1.7\textunderscore 10yrs} \texttt{OpSim} run. We show both cadences including two 15 s exposures per visit and cadences with single 30 s exposures. Among the former, none seems more favourable than the \texttt{baseline} simulation, overall. The single 30 s exposure cadences lead to some time saving and so to more observations. Apart from the two extreme \texttt{filterdist} simulations, which strongly favour either blue or red filters, with super sampling in these filters and undersampling in the other bands, the other \texttt{OpSim} runs can provide some more visits than the \texttt{baseline} in some filters, but in general the difference is much less than 50 visits in 10 yr. This small advantage would occur at the expenses of doubling the number of saturated observations. In particular, the \texttt{baseline\textunderscore nexp1\textunderscore v1.7\textunderscore 10yrs} yields -2,      -3,      10,      20,      18,     -11 data points in the $u,g,r,i,z,y$ bands in 10 years with respect to the \texttt{baseline\textunderscore nexp2\textunderscore v1.7\textunderscore 10yrs}, which is a negligible difference.

\begin{figure}[ht!]
\plottwo{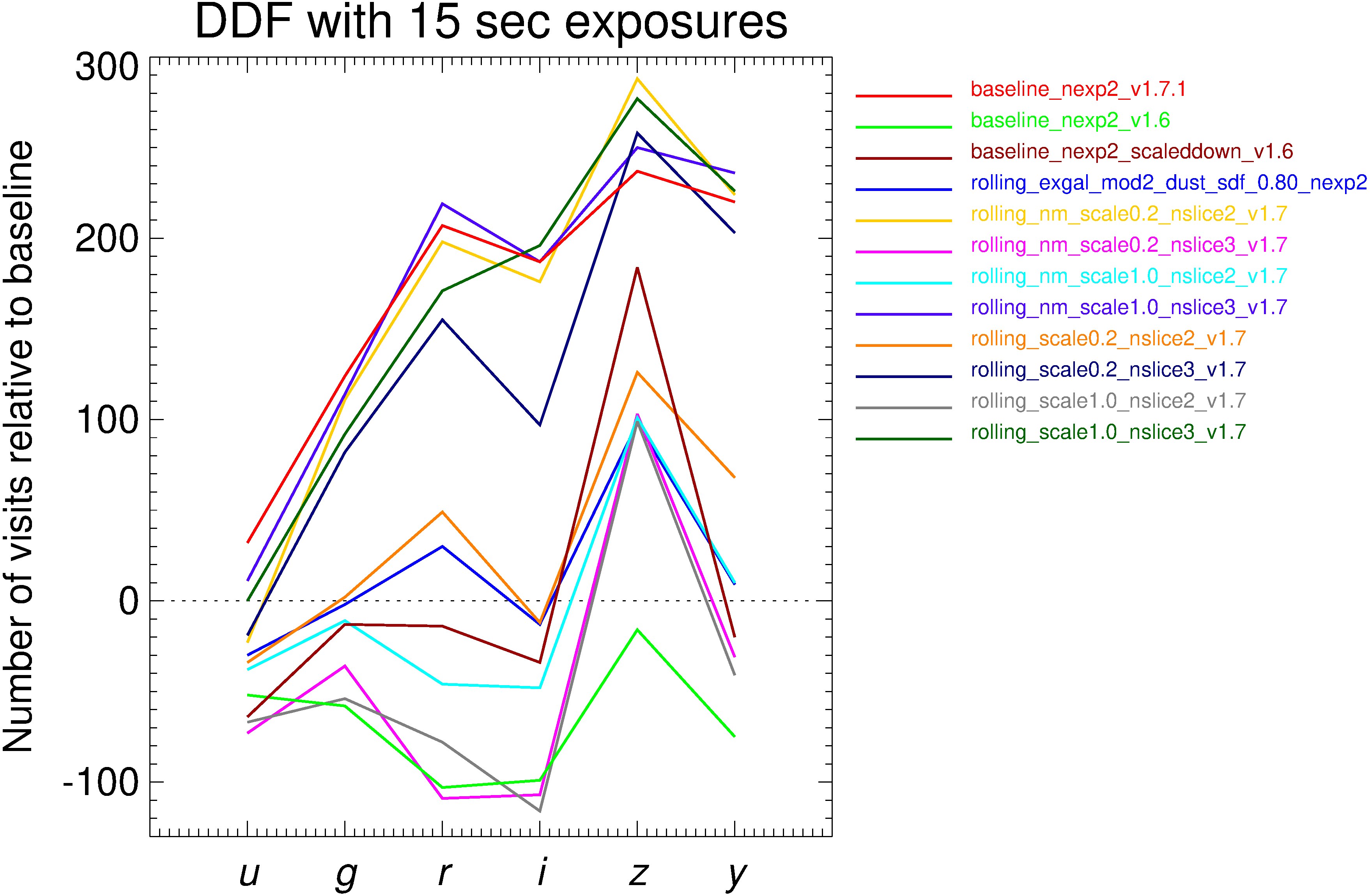}{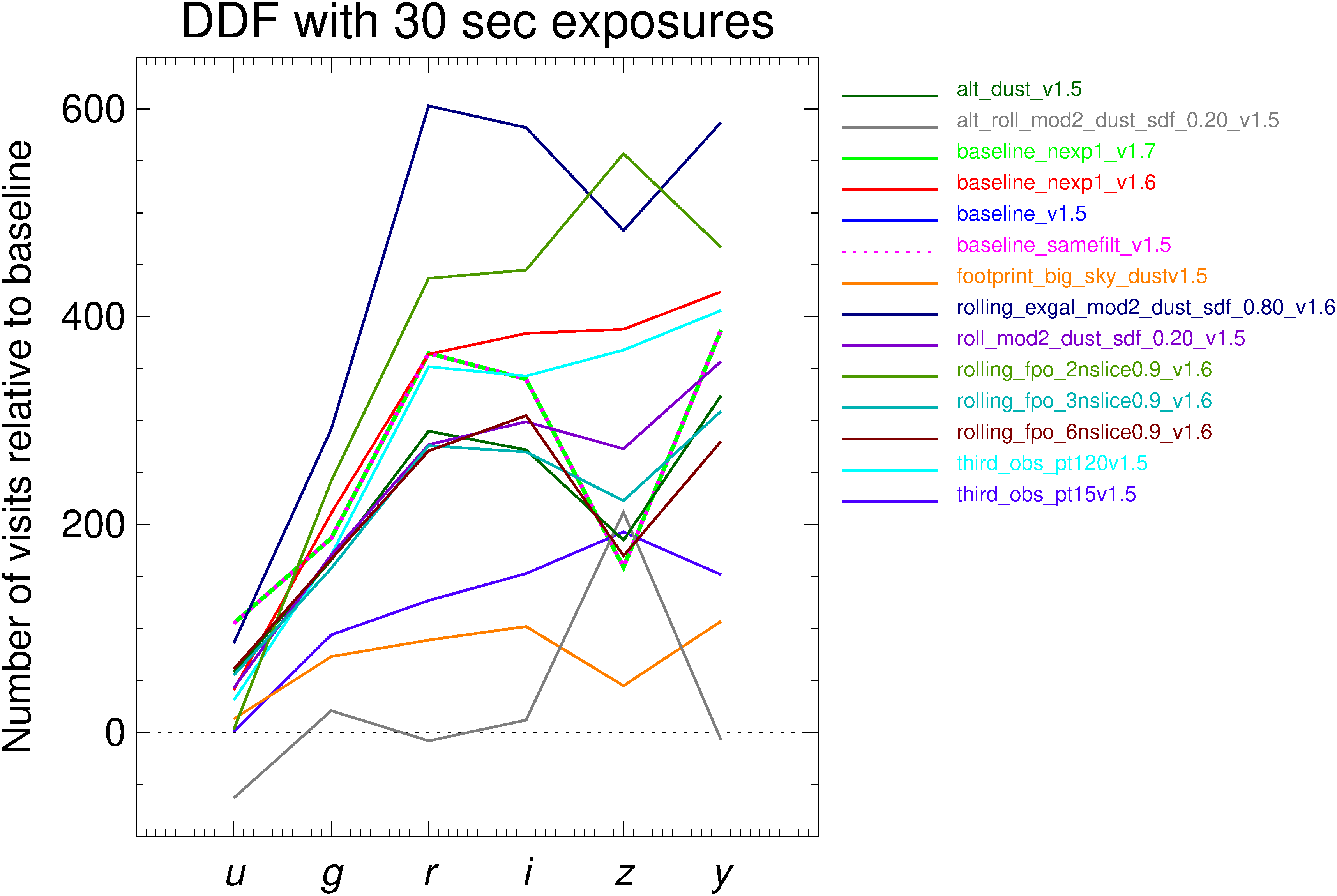}
\caption{Difference between the number of DDF visits in various cadence simulations and that of the \texttt{baseline\textunderscore nexp2\textunderscore v1.7\textunderscore 10yrs} cadence. \texttt{OpSim} runs with either double 15 s (left) or single 30 s (right) exposure times are shown.  \label{fig:filters_ddf}}
\end{figure}

The situation changes dramatically in the case of DDF, as shown in Figure \ref{fig:filters_ddf}. 
Some rolling cadence simulations with two 15 s exposures already imply many more observations than the corresponding \texttt{baseline} cadence. The number further increases with single 30 s exposures.
The \texttt{OpSim} run \texttt{baseline\textunderscore nexp1\textunderscore v1.7\textunderscore 10yrs} provides 105,     187,     365,     340,     159, and 387 more data points in the $u,g,r,i,z,y$ filters, respectively. 
The \texttt{rolling\textunderscore extragalactic} cadence appears as an extreme case, yielding 86,     292,     603,     582,     483, and      587 more visits.
As detailed by \citet{jones2020}, this cadence concentrates on low-extinction regions and divides the sky in quarters, so that  there is always a region of observing emphasis that is reachable by northern telescopes.
However, the increase of 603 data points in the $r$ band corresponds to an increase of only 22 nights in 10 yr, which is not a great gain, when considering that it would worsen the saturation problem.

Actually, there is a series of \texttt{OpSim} runs specifically testing different DDF cadences \citep{jones2020}.
In particular, the \texttt{daily\textunderscore ddf\textunderscore v1.5\textunderscore 10yrs} observing strategy leads to a much larger number of nights, as can be seen in Table \ref{tab:lcsamp}.
However, even in this case, the inter-day and intra-day samplings are not optimal.
This is shown in Figure \ref{fig:dt}, where the time difference between subsequent observations in the \texttt{daily} cadence can be compared to that in the \texttt{baseline}.
Although the \texttt{daily} cadence leads to a better sampling of the inter-day time-scales and to a slightly better sampling of the intra-day time-scales, a large percentage of all observation pairs (72\% for the $i$ and $y$ bands and 64\% for the $r$ and $z$ bands) still has a time separation within 1 minute.

\section{Discussion and conclusions}
In this paper we have analysed the impact of different choices for the Rubin-LSST observing strategy on the study of both brightness and spectral blazar variability.
We first discussed the number of blazar candidates that Rubin-LSST will likely see and found that it is of the order of several tens of thousands. Most of them should be detectable with a single-visit exposure, making a multiband variability study possible. This estimate is obtained both from general considerations and by taking the largest catalogue of blazar candidates into account. If it is correct, besides the WFD, also the DDFs will be of interest for blazar variability studies, as they will contain more than a hundred objects on which a more detailed investigation of the flux and spectral variability will be possible.

Saturation is a serious problem, since it could affect the most interesting events (outbursts), which are eligible of multiwavelength follow-up observations. 
Saturation will also affect the brightest blazars,  many of which are alleged to be sources of high-energy neutrinos. 
In the white paper by \citet{raiteri2018}, we suggested a minisurvey in star trail mode, which could solve the problem of blazar saturation and at the same time offer the opportunity to search for very rapid transients, like possible optical counterparts of fast radio bursts. We are aware that this idea may present technical problems and may be expensive in terms of time. Another possibility to overcome the saturation problem, which was proposed in the same paper, would be to diversify the two exposure times in a single visit, using e.g.\ 2+28 sec or 5+25 sec snapshots.
Among the available \texttt{OpSim} runs, the number of saturated observations in a simulation with reasonable flaring conditions, goes from about 3\% to about 6\%, when we compare cadences with two exposures of 15 s per visit to those with single 30 s exposures.
Some cadences with single 30 s exposures provide much more observations, especially in DDFs, but only a few more observing nights. Indeed, DDFs are characterized by sequences of intranight observations in the same filter, but these sequences are so short, about 11 min, that they cannot in general trace the fastest flux and spectral changes, even in OVV sources. 

As mentioned, the understanding of blazar variability requires intensive multiwavelength monitoring. In case of outstanding flaring events, follow-up observations should be triggered, especially at X-ray and $\gamma$-ray energies. Indeed, the high-energy counterparts of optical flares usually occur almost contemporaneously, so that prompt alerts are required.
This can be accomplished through the Rubin-LSST brokers \citep{bellm2020}, which
are expected to release millions of alerts per night on objects that can be detected with at least $5 \sigma$ on the difference images. 
Many of these alerts will concern blazars.
Beneficial follow-up observations should also be carried out in polarimetric mode. 
Indeed, blazar synchrotron emission is polarized, and both the polarization degree and angle can be quite variable. 
The polarimetric behaviour gives information on the magnetic field in the jet, but also on the jet geometrical structure \citep[e.g.][]{larionov2013,raiteri2021_galaxies}.

Our recommendations for the  choice of the Rubin-LSST cadence from the point of view of blazar variability are:
\begin{itemize}
\item In the WFD, double 15 s exposures are favoured with respect to single 30 s ones, to mitigate saturation of flaring/bright blazars. 
No cadence was found to perform much better than the \texttt{OpSim} run \texttt{baseline\textunderscore nexp2\textunderscore v1.7\textunderscore 10yrs}, whose major drawback is the coupling of adjacent filters in the same night (see below).
\item In DDFs, the problem of saturation should have a limited impact, and cadences with single 30 s exposures providing much more sampling can be considered. However, a more beneficial cadence should include shorter DDF exposure sequences more often, and in particular intranight observations of the same DDF with more time spacing between visits in the same filter. In this sense, a DDF observing strategy like that included in the \texttt{daily\textunderscore ddf\textunderscore v1.5\textunderscore 10yrs} \texttt{OpSim} run represents a better choice than the \texttt{baseline} implementation, but this can be further improved introducing a more homogeneous sampling of the different variability time-scales  \citep[see also][]{bellm2021}.
\item In both WFD and DDFs, visits with different filters in the same night are required to obtain colour indices with data close in time, which is necessary to avoid bias in colours due to variability; this means that the \texttt{samefilt} \texttt{OpSim} runs are detrimental.
\item Colour variability and trends with brightness are clearer when colours are obtained with data in filters that are more separated in wavelength; therefore, in both WFD and DDFs the choice of filters to couple in the same night should prefer bands that are not contiguous. 
\item We favour an extension of the WFD footprint to the North, with an enlargement of the low-extinction extragalactic sky, which would also be reachable by many more observing facilities that could complement the Rubin-LSST monitoring. This should be accomplished without decreasing the sampling.
\item We cannot support a distribution of filters more skewed towards blue/red filters than in the \texttt{baseline} cadence, because different types of blazars would require different choices of filters.
\item Observations at high air masses may be beneficial, as they can prolong the observing season and lead to smaller gaps in the light curves. Moreover, high air mass can also avoid saturation in some cases.
\item ToO observations should be considered when high-energy neutrinos of astrophysical origin are detected by neutrino facilities. The uncertainty on the arrival direction can be up to some degrees. Since the most promising candidates as neutrino sources are likely bright blazars, the ToO observations should include short exposures to avoid possible saturation.

\end{itemize}

We finally mention that besides variability, Rubin-LSST will also improve other aspects of our knowledge of blazars, like census and environment.
Rubin-LSST will potentially identify thousands of new objects. These will likely be FSRQs at high redshift, with important cosmological implications.
As already mentioned, blazars belong to the radio-loud AGN population, which seems to require SMBH with masses greater than $10^8 \, M_\sun$. Blazars at high redshift thus imply the presence of SMBH in the early Universe. Up to now, the blazar with the highest redshift ($z>6$) has been discovered by \citet{belladitta2020}. 
The identification of new blazar candidates with Rubin-LSST must rely on a combination of colours and variability features, and possibly on multiwavelength information, also thanks to new observing facilities with enhanced capabilities.
In the X-ray band, the extended ROentgen Survey with an
Imaging Telescope Array \citep[{\em eROSITA};][]{predehl2021} has been performing an all-sky survey since 2019, with a sensitivity of $10^{-14} \rm \, erg \, cm^{-2} \, s^{-1}$. The survey will include eight scans of the whole sky in four years.
The Advanced Telescope for High ENergy Astrophysics \citep[{\em Athena};][]{barcons2017}, to be launched in 2030, will perform various surveys, further improving depth. 
In particular, the ``wide" survey will cover 48 deg$^{2}$ and include the Rubin-LSST DDFs, with a sensitivity of $\sim 10^{-16} \rm \, erg \, cm^{-2} \, s^{-1}$ in the 0.5--2.0 keV energy range.
For comparison, the 0.1--2.4 keV flux of the 2249 blazars in BZCAT5 for which X-ray data are available,
ranges between $2 \times 10^{-14}$ and $3.2 \times 10^{-10} \rm \, erg \, cm^{-2} \, s^{-1}$. 
An ultra-wide ($\sim 800 \rm \, deg^{2}$) and shallow ($\sim 5 \times 10^{-16} \rm \, erg \, cm^{-2} \, s^{-1}$) {\em Athena} survey, which would nicely complement Rubin-LSST, although still on a smaller area, is under consideration.
At $\gamma$ rays, the {\em Fermi} satellite \citep{atwood2009} has been scanning the entire sky each day since 2008. The {\em Fermi} Large Area Telescope (LAT) fourth source catalog \citep[4FGL;][]{abdollahi2020} includes 5064 sources, $\sim 62\%$ of which are blazars, but there are still 1336 unassociated sources, most of which are expected to be blazars, for which Rubin-LSST may find the optical counterpart.
Finally, there are 80 blazars detected at very high energies (VHE; E $>$ 100 GeV) by current Cherenkov telescopes\footnote{From the TeVCat catalogue, http://tevcat.uchicago.edu/}. 
Fifty of them lie at $-90^{\circ} < \delta < +30^{\circ}$ and are likely observable by Rubin-LSST. According to BZCAT5, their magnitude is  $11 < R < 19.4$, with 25 sources brighter than 15.5 mag, and thus possibly affected by saturation. 
Moreover, the second catalog of hard {\em Fermi}-LAT sources
\citep[2FHL;][]{ackermann2016} reports the detection of 265 blazars in
the 50 GeV--2 TeV energy range, most of them potentially detectable by the next-generation Cherenkov Telescope Array \citep[CTA;][]{acharya2019}, and whose optical counterpart is expected to be identifiable by Rubin-LSST.
The blazar candidates will then need validation through follow-up spectroscopic observations.

Finally, the depth of Rubin-LSST observations will allow us to study the blazar environment, which in turn is a fundamental ingredient to understand what is the parent population of blazars among the unbeamed AGN sources. 
\citet{muriel2016} found that a high percentage of BL Lac-type objects lies in group of galaxies.
However, \citet{sandrinelli2019} claimed that the environment of BL Lac-type sources is by a factor two less rich than that of Fanaroff-Riley type I (FR I) radio galaxies, which represent their parent population according to the AGN unification scheme. This result, which questions the identification of the parent population of the BL Lac-type objects, needs to be verified with larger and more uniform  samples of FR I and BL Lac-type objects.

\begin{acknowledgments}
We are deeply grateful to Rachel Street, Federica Bianco and Niel Brandt for their essential coordination work inside the Rubin LSST Science Collaborations.
This paper was created in the nursery of the Rubin LSST ``Transients and Variable Stars" Science Collaboration ({\url{https://lsst-tvssc.github.io/}}).
The authors acknowledge the support of the Vera C. Rubin Legacy Survey of Space and Time ``Transients and Variable Stars" and ``AGN" Science Collaborations that provided opportunities for collaboration and exchange of ideas and knowledge.
The authors are thankful for the support provided by the Vera C. Rubin Observatory \texttt{MAF} team in the creation and implementation of \texttt{MAF}s.
The authors acknowledge the support of the LSST Corporation, which enabled the organization of many workshops and hackathons throughout the cadence optimization process.

The authors acknowledge the use of data taken and assembled by the WEBT collaboration and stored in the WEBT archive at the Osservatorio Astrofisico di Torino - INAF (http://www.oato.inaf.it/blazars/webt/).
\end{acknowledgments}

%

\vspace{5mm}
\facilities{Vera C. Rubin Observatory}


\software{LSST metrics analysis framework \citep[\texttt{MAF};][]{jones2014}
          }


\bibliography{blazars4apjs}{}
\bibliographystyle{aasjournal}



\end{document}